\documentclass[envcountsame,orivec,fleqn]{llncs}

\usepackage[hyphens]{url}
\usepackage{cite}
\usepackage{hyperref}
\usepackage{amsmath}
\usepackage[english]{babel}
\usepackage{latexsym}
\usepackage{color}
\usepackage[usenames,dvipsnames]{xcolor}
\usepackage{amssymb}	\usepackage{verbatim}   \usepackage{ifthen}
\usepackage{xfrac}      \usepackage{graphicx}   \usepackage{wrapfig}
\usepackage[plain]{algorithm}
\usepackage{algpseudocode}
\usepackage{paralist}
\usepackage{array}
\usepackage{multirow}
\usepackage{rotating}
\usepackage[capitalise]{cleveref}
\hypersetup{breaklinks=true}

\makeatletter
\newlength{\@auxiliaryquotewidth}

\newenvironment{labelledquote}
               {\setlength{\@auxiliaryquotewidth}{\textwidth}\addtolength{\@auxiliaryquotewidth}{-\leftmargin}\list{}{\rightmargin1.5\leftmargin}                \item\relax\refstepcounter{equation}\makebox[-4pt][l]{\makebox[\@auxiliaryquotewidth][r]{(\theequation)}}}               {\endlist}

\makeatother

\Crefname{equation}{}{}
\Crefname{proposition}{Prop.}{Props.}

\renewcommand{\emptyset}{\varnothing}

\newcommand{\Reals}{\ensuremath{\mathbb{R}}}

\newcommand{\N}{\ensuremath{\mathbb{N}}}

\newcommand{\NOx}{\ensuremath{\mathrm{NO}_x}}

\newcommand{\Parameters}{{\mathsf{Param}}}
\newcommand{\ParamofInt}{{\mathsf{PIntrs}}}
\newcommand{\Inputs}{{\mathsf{In}}}
\newcommand{\Outputs}{{\mathsf{Out}}}

\newcommand{\inp}{\mathsf{i}}
\newcommand{\outp}{\mathsf{o}}
\newcommand{\param}{\mathsf{p}}

\newcommand{\Norm}{{\mathsf{StdIn}}}
\newcommand{\Commit}{{\mathsf{Comm}}}

\newcommand{\inpbound}{\kappa_\inp}
\newcommand{\outpbound}{\kappa_\outp}

\newcommand{\robustly}{robustly}

\newcommand{\Hausdorff}{\mathcal{H}}

\newcommand{\Vars}{\mathsf{Var}}
\newcommand{\Vals}{\mathsf{Val}}
\newcommand{\wpre}{\ensuremath{\mathop{\mathrm{wp}}}}

\newcommand{\lImp}{\Rightarrow}

\newcommand{\limp}{\rightarrow}
\newcommand{\liff}{\leftrightarrow}
\newcommand{\true}{\textsf{true}}

\makeatletter
\newcommand{\dropthen}[1][0]{\\\hspace{\algorithmicindent}\hspace{-.2em}\hspace{\dimexpr 1.3em * #1}}
\makeatother
\newcommand{\algorithmsize}{\fontsize{8pt}{9.6pt}\selectfont}
\newcommand{\var}[1]{\ensuremath{\textit{#1}}}
\newcommand{\const}[1]{\ensuremath{\textsf{#1}}}

\newcommand{\abs}[1]{\lvert {#1} \rvert}
\newcommand{\last}{\mathop{\mathrm{last}}}

\newcommand{\ap}{\mathsf{AP}}
\newcommand{\X}{\mathop{\mathsf{X}}}
\newcommand{\F}{\mathop{\mathsf{F}}}
\newcommand{\G}{\mathop{\mathsf{G}}}
\newcommand{\U}{\mathbin{\mathsf{U}}}
\newcommand{\W}{\mathbin{\mathsf{W}}}
\newcommand{\E}[1]{\mathop{\exists{#1}.}}
\newcommand{\A}[1]{\mathop{\forall{#1}.}}

\newcommand{\tr}{t}
\newcommand{\tass}{\Pi}
\newcommand{\sat}[1]{\mathrel{\models_{#1}}}

\definecolor{lightblue}{RGB}{210,210,225}
\definecolor{lightred}{RGB}{225,210,210}
\definecolor{lightgreen}{RGB}{210,225,210}
\definecolor{lightyellow}{RGB}{225,222,200}
\definecolor{lightpurple}{RGB}{225,210,225}
\definecolor{darkergreen}{RGB}{0,64,0}
\definecolor{darkred}{RGB}{128,0,0}
\definecolor{darkblue}{RGB}{0,0,128}
\definecolor{darkgreen}{RGB}{0,128,0}
\definecolor{darkpurple}{RGB}{128,0,128}

\newcommand{\colorpar}[3]{\colorbox{#1}{\parbox{#2}{#3}}}

\newcommand{\marginremark}[4]{\marginpar{\colorpar{#2}{\linewidth}{\color{#1}\tiny{[#3]~ #4}}}}

\newcommand{\citeref}[2]{\href{#1}{#2}~\cite{#1}}

\makeatletter
\def\THICKhrulefill{\leavevmode \leaders \hrule height 5pt\hfill \kern \z@}
\makeatother

\iftrue
  \newcommand{\remarkBF}[1]{\marginremark{darkpurple}{lightpurple}{BF}{#1}}
  \newcommand{\remarkGB}[1]{\marginremark{darkgreen}{lightyellow}{GB}{#1}}
  \newcommand{\remarkHH}[1]{\marginremark{darkblue}{lightblue}{HH}{#1}}
  \newcommand{\remarkPRD}[1]{\marginremark{darkred}{lightred}{PRD}{#1}}
  \newcommand{\remarkSB}[1]{\marginremark{darkergreen}{lightgreen}{SB}{#1}}
\else
  \newcommand{\remarkPRD}[1]{}
  \newcommand{\remarkGB}[1]{}
  \newcommand{\remarkSB}[1]{}
  \newcommand{\remarkBF}[1]{}
  \newcommand{\remarkHH}[1]{}
\fi

\titlerunning{Is your software on dope? Formal analysis of surreptitiously ``enhanced'' programs}
\authorrunning{P.R.~D'Argenio \and G.~Barthe \and S.~Biewer \and B.~Finkbeiner \and H.~Hermanns}

\begin{document}
\title{
  Is your software on dope?  \thanks{This work is partly supported by the ERC Grants 683300 (OSARES) and 695614 (POWVER), by the Saarbr\"ucken Graduate School of Computer Science, by the Sino-German CDZ project 1023 (CAP), by ANPCyT PICT-2012-1823, by SeCyT-UNC 05/BP12 and 05/B497, and by the Madrid Region project S2013/ICE-2731 N-GREENS Software-CM.	}}
\subtitle{
  Formal analysis of surreptitiously ``enhanced'' programs}
\author{
	Pedro R.~D'Argenio\inst{1,2} \and 
	Gilles Barthe\inst{3}    \and
	Sebastian Biewer\inst{2} \and\\
	Bernd Finkbeiner\inst{2} \and
	Holger Hermanns\inst{2}
}\institute{
FaMAF, Universidad Nacional de C\'ordoba -- CONICET
    \and 
  Saarland University -- Computer Science, Saarland Informatics Campus
    \and
  IMDEA Software
   }
\maketitle

\begin{abstract}
Usually, it is the software manufacturer who employs verification or
testing to ensure that the software embedded in a device meets its
main objectives.  However, these days we are confronted with the
situation that economical or technological reasons might make a
manufacturer become interested in the software slightly deviating from
its main objective for dubious reasons.
Examples include lock-in strategies and the \NOx\ emission scandals in
automotive industry.
This phenomenon is what we
call \emph{software doping}. It is turning more widespread as software
is embedded in ever more devices of daily use.

The primary contribution of this article is to provide a hierarchy of
simple but solid formal definitions that enable to distinguish whether
a program is \emph{clean} or \emph{doped}.  Moreover, we show that
these characterisations provide an immediate framework for analysis by
using already existing verification techniques.  We exemplify this by
applying self-composition on sequential programs and model checking of
HyperLTL formulas on reactive models.
\end{abstract}

\section{Introduction}

The \citeref{https://en.wikipedia.org/wiki/Volkswagen_emissions_scandal}{Volkswagen exhaust emissions scandal} has put \emph{software doping} in the spotlight:  
Proprietary embedded control software does not always exploit functionality offered by a device in the best interest of the device owner. Instead the software may be tweaked in various manners,
driven by interests different from those of the owner or of society. This is indeed a common characteristics for the manner how different manufacturers~\citeref{http://www.reuters.com/article/us-fiat-emissions-germany-idUSKCN0XL0MT,http://www.reuters.com/article/us-volkswagen-emissions-germany-opel-idUSKCN0Y92GI} {circumvented} the diesel emission regulations around the world.  The exhaust software was manufactured in such a way
that it heavily polluted the environment, unless the software detected
the car to be (likely) fixed  on a particular test setup used to determine the
\NOx\ footprint data officially published. Phenomena resembling the emission scandal have also been reported in the context of
\citeref{http://forums.appleinsider.com/discussion/158782/galaxy-s-4-on-steroids-samsung-caught-doping-in-benchmarks}{smart
  phone designs}, where software was tailored to perform better when
detecting it was running a certain benchmark, and otherwise running in
lower clock speed. Another smart phone case,
\citeref{https://www.theguardian.com/money/2016/feb/05/error-53-apple-iphone-software-update-handset-worthless-third-party-repair}{disabling
  the phone} via a software update after ``non-authorised'' repair,
has later been
\citeref{https://techcrunch.com/2016/02/18/apple-apologizes-and-updates-ios-to-restore-iphones-disabled-by-error-53/}{undone}.

Usually, it is the software manufacturer who employs verification or testing to ensure that the software embedded in a device meets its main objectives.  However, these days we
are confronted with the situation that economical or technological reasons might make a  manufacturer become interested in the software slightly deviating from its main objective for  dubious reasons.  This phenomenon is what we call
\emph{software doping}. It is turning more widespread as software is
embedded in ever more devices of daily use.

The simplest and likely most common example of software doping (effectuating  a customer lock-in strategy~\cite{ecojournal1989}) is that
of  \citeref{http://www.wasteink.co.uk/epson-firmware-update-compatible-problem/}{ink printers} refusing to work when supplied with a toner or ink cartridge of a \citeref{https://conversation.which.co.uk/technology/printer-software-update-third-party-printer-ink/}{third party manufacturer}, albeit being technically compatible. Similarly, cases are known where \citeref{https://nctritech.wordpress.com/2010/01/26/dell-laptops-reject-third-party-batteries-and-ac-adapterschargers-hardware-vendor-lock-in/}{laptops refuse to charge} the battery if connected to a third-party charger. 
 More subtle variations of this kind of doping just issue a warning
message about the \citeref{http://uk.pcmag.com/printers/60628/opinion/the-secret-printer-companies-are-keeping-from-you}{risk of using a ``foreign'' cartridge}. In the same
vein, it is known that printers \citeref{http://www.slate.com/articles/technology/technology/2008/08/take_that_stupid_printer.html}{emit ``low toner'' warnings} earlier than needed, so as to drive or force the customer into replacing cartridges prematurely. Moreover, there are allegations that
\citeref{http://www.mintpressnews.com/214505-2/214505/}{software
  doping has occurred in the context of electronic-voting so as to
  manipulate the outcome}.  Tampering with voting machines has been
proved a relatively easy task~\cite{FeldmanHF07:uss}.
Common to all these examples is that the software user has little or no control over its execution, and that the functionality in
question is against the interests of user or of society.

Despite the apparently pervasive presence of software doping, a systematic investigation or formalisation  from the software engineering perspective is not existing. Fragmentary 
attention has been payed in the security domain with respect to cryptographic
protections being sabotaged by insiders~\cite{SchneierFKR15}.
Typical examples are the many known backdoors, including the prominent dual EC deterministic random bit generator standardised by NIST~\cite{CheckowayNE0LRBMSF14:uss}. Software doping however goes far beyond inclusion of backdoors.

Despite the many examples, it is not at all easy to provide a crisp characterisation of what
constitutes software doping. This paper explores this issue, and proposes a hierarchy of  formal characterisations of software doping. We aim at formulating and enforcing
rigid requirements on embedded software driven by public interest, so
as to effectively ban software doping.  In order to sharpen our
intuition, we offer the following initial characterisation attempt~\cite{BartheDFH16:isola}.

\begin{labelledquote}\label{dope:intuition}
  A software system is doped if the manufacturer has included a hidden
  functionality in such a way that the resulting behaviour
  intentionally favors a designated party, against the interest of
  society or of the software licensee.
\end{labelledquote}

So, a doped software induces behaviour that can not be justified by
the interest of the licensee or of society, but instead serves another
usually hidden interest. It thereby favors a certain brand, vendor,
manufacturer, or other market participant. This happens intentionally,
and not by accident.
However, the question whether a certain behaviour is intentional or
not is very difficult to decide. To illustrate this, we recall that
the above mentioned smart phone case, to be specific the iPhone-6, where
\citeref{https://www.theguardian.com/money/2016/feb/05/error-53-apple-iphone-software-update-handset-worthless-third-party-repair}{``non-authorised''
  repair rendered the phone unusable} after an iOS update, seemed to
be intentional when it surfaced, but was actually tracked down to a
software glitch of the update and fixed later. Notably, if the iOS
designers would have had the particular intention to mistreat
licensees who went elsewhere for repair, the same behaviour could well
have qualified as software doping in the above
sense~\cref{dope:intuition}.
As a result, we will look at software doping according to the above
characterisation, keeping in mind the possibility of intentionality
but not aiming to capture it in a precise manner.

In our work, we  use concise examples that are directly inspired by the real cases  reviewed above.  They motivate our hierarchy of formal characterisations of \emph{clean} or \emph{doping-free}
software.

A core observation will be that software doping can be characterised by considering 
the program if  started  from two different but compatible initial states. If the obtained outputs are not compatible, then this implies that the software is \emph{doped}.  Thinking in terms of the printer, one would expect that printing with different but compatible
cartridges would yield the same printout without any alteration in the observed 
alerts. As a consequence, the essence of the property of being clean can be cast as a
\emph{hyperproperty}~\cite{ClarksonS08,ClarksonS10:jcs}.

We first explore characterisations on sequential software
(\cref{sec:sd:seq}).  We introduce a characterisation that ensures the
proper functioning of the system whenever it is confined to standard
parameters and inputs. Afterwards, we give two other characterisations
that limit the behaviour of the system whenever it goes beyond
such standard framework.  We then revise these characterisations so as to apply to 
reactive non-deterministic systems (\cref{sec:sd:react}).

Traditionally hyperproperties require to be analysed in an
\emph{ad-hoc} manner depending on the particular property.  However,
a  general framework is provided by techniques based on, e.g.,
self-composition techniques~\cite{BartheDR04:csfw} or specific logic
 such as HyperLTL~\cite{ClarksonFKMRS14:post}.
 Indeed, we show (\cref{sec:sd:seq:selfcomp}) how these properties can be analysed using self-composition on deterministic programs, particularly using weakest precondition reasoning~\cite{book:dijsktra}, and  we do the same (\cref{sec:sd:react:hltl}) for reactive systems using HyperLTL.  In both settings we demonstrate principal feasibility by presenting verification studies  of simple but representative examples.

\section{Software Doping on Sequential Programs}
\label{sec:sd:seq}

Think of a program as a function that accepts some initial parameters
and, given some inputs, produces some outputs, maybe in a
non-deterministic manner.
Thus, a parameterised sequential non-deterministic program is a
function $S:\Parameters\to\Inputs\to 2^\Outputs$, where $\Parameters$
is a set of parameters, each one of them fixing a particular instance
of the program $S$, and $\Inputs$ and $\Outputs$ being respectively
the sets of inputs accepted by $S$ and outputs produced by $S$.
Notice that for a fixed parameter $\param$ and input $\inp\in\Inputs$,
the run of program $S(\param)(\inp)$ may give a set of possible
outputs.

\begin{figure}[t]
\begin{minipage}{0.48\textwidth}
  \centering\algorithmsize  \begin{algorithmic}
  \Procedure{Printer}{\var{cartridge\_info}}
  \If {$\Call{type}{\var{cartridge\_info}}\in\const{Compatible}$\dropthen}
    \State \Call{read}{\var{document}}
    \State \Call{print}{\var{stdout},\var{document}}
              \Else
    \State \Call{turnOn}{\var{alert\_led}}
  \EndIf
  \EndProcedure
  \end{algorithmic}
  \caption{A simple printer.}\label{fig:printer:general}
\end{minipage}\hfill
\begin{minipage}{0.48\textwidth}
  \centering\algorithmsize  \begin{algorithmic}
  \Procedure{Printer}{\var{cartridge\_info}}
  \If {$\Call{brand}{\var{cartridge\_info}} = \var{my-brand}$\dropthen}
    \State \Call{read}{\var{document}}
    \State \Call{print}{\var{stdout},\var{document}}
              \Else
    \State \Call{turnOn}{\var{alert\_led}}
  \EndIf
  \EndProcedure
  \end{algorithmic}
  \caption{A doped printer.}\label{fig:printer:doped}
\end{minipage}
\end{figure}
To understand a first possible definition, consider the program
embedded in a printer (a simple abstraction is given in
\cref{fig:printer:general}).  This program may check compatibility
of the ink or toner cartridge and print whenever the cartridge is
compatible.  In this case, we can think of the program
\textsc{Printer} as a function parameterised with the information on
the cartridge, that receives a document as input and produces a
sequence of pages as outputs whenever the cartridge is compatible,
otherwise it turns on an alert led.  In this setting, we expect that the
printer shows the same input-output behaviour for any compatible
cartridge.

A printer manufacturer may manipulate this program in order to favour
its own cartridge brand.  An obvious way is displayed in
\cref{fig:printer:doped}.  This is a sort of discrimination based on
parameter values.
Therefore, a first approach to characterising a program as
\emph{clean} (or \emph{doping-free}) is that it should behave in a
similar way for all parameters of interest.  By ``similar behaviour''
we mean that the visible output should be the same for any given input
in two different instances of the same (parameterised) program.  Also,
by ``all parameters of interest'', we refer to all parameter values we
are interested in.  In the case of the printer, we expect that it
works with any \emph{compatible} cartridge, but not with every
cartridge.  Such a compatibility domain defines a first scope within
which a software is evaluated to be clean or doped.

Formally, if $\ParamofInt\subseteq\Parameters$, we could say that a
parameterised program $S$ is clean (or doping-free) if
for all pairs of parameters of interest $\param,\param'\in\ParamofInt$
and input $\inp\in\Inputs$, $S(\param)(\inp) = S(\param')(\inp)$.
Thus, the program of \cref{fig:printer:general} satisfies this
constraint whenever \const{Compatible}\ is the set of parameters of
interest (i.e.\ $\const{Compatible}=\ParamofInt$).  Instead, the
program of \cref{fig:printer:doped} would be rejected as \emph{doped}
by the previous definition.

We could imagine, nonetheless, that the printer manufacturer may like
to provide extra functionalities for its own product which is outside
of the standard for compatibility.  For instance (and for the sake of
this discussion) suppose the printer manufacturer develops a new file
format that is more efficient or versatile at the time of printing,
but this requires some new technology on the cartridge (we could
compare this to the introduction of the postscript language when
standard printing was based on dots or ASCII code).
The manufacturer still wants to provide
the usual functionality for standard file formats that work with
standard compatible cartridges and comes up with the program of
\cref{fig:printer:nondoped}.
Notice that this program does not conform to the specification of a
clean program as given above since it behaves differently when a
document of the new (non-standard) type is given.  This is clearly not
in the spirit of the program in \cref{fig:printer:nondoped} which is
actually conforming to the expected requirements.

\begin{wrapfigure}[14]{r}{6.8cm}
  \algorithmsize\vspace{-2em}  \begin{algorithmic}
    \Procedure{Printer}{\var{cartridge\_info}}
    \If {$\Call{type}{\var{cartridge\_info}}\in\const{Compatible}$}
      \State \Call{read}{\var{document}}
      \If {$({\neg \Call{newType}{\var{document}}}{}$\\
           \hspace{4em} ${}\lor {}{\Call{supportsNewType}{\var{cartridge\_info}}})$\dropthen[1]}
        \State \Call{print}{\var{stdout},\var{document}}
                              \Else
        \State \Call{turnOn}{\var{alert\_led}}
      \EndIf
    \Else
      \State \Call{turnOn}{\var{alert\_signal}}
    \EndIf
    \EndProcedure
  \end{algorithmic}
  \caption{A clean printer.}\label{fig:printer:nondoped}
\end{wrapfigure}Thus, our first definition states that a program is \emph{clean} if,
for any possible instance from the set of parameters of interest, it
exhibits the same visible outputs when supplied with the same input,
provided this input complies with a given standard.
Formally, we assume a set $\ParamofInt\subseteq\Parameters$ of
parameters of interest and a set $\Norm\subseteq\Inputs$ of standard
inputs and propose the following definition.

\begin{definition}\label{def:clean:seq}
  A parameterised program $S$ is \emph{clean} (or \emph{doping-free})
  if for all pairs of parameters of interest
  $\param,\param'\in\ParamofInt$ and input $\inp\in\Inputs$, if
  $\inp\in\Norm$ then $S(\param)(\inp) = S(\param')(\inp)$.
    If the program is not clean we will say that it is \emph{doped}.
\end{definition}

The characterisation given above is based on a comparison of the
behaviour of two instances of a program, each of them responding to
different parameter values within $\ParamofInt$.  A second, different
characterisation may instead require to compare a reference
specification capturing the essence of clean behaviour against any
possible instance of the program.  The first approach seems more
general than the second one in the sense that the specification could
be considered as one of the possible instances of the (parameterised)
program.
However, we can consider a distinguished parameter $\hat{\param}$ so
that the instance $S(\hat{\param})$ is actually the specification of
the program, in which case, both definitions turn out to be equivalent.
In any case, it is important to observe that the specification may not
be available since it is also made by the software manufacturer, and
only the expected requirements may be known.

We remark that \cref{def:clean:seq} entails the existence of a
contract which defines the set of parameters of interest and the set
of standard inputs.
In fact, \cref{def:clean:seq} only asserts doping-freedom if the
program is well-behaved within such a contract, namely, as long as the
parameters are within $\ParamofInt$ and inputs are within $\Norm$.  A
behaviour outside this realm is deemed immediately correct since it is
of no interest.
This view results too mild in some cases where the change of behaviour
of a program between a standard input and a non-standard but yet
not-so-different input is extreme.

\begin{wrapfigure}[7]{r}{5.1cm}
  \algorithmsize  \begin{algorithmic}
    \Procedure{EmissionControl}{{}}
      \State \Call{read}{\var{throttle}}
      \State \var{def\_dose} := \Call{SCRModel}{\var{throttle}}
      \State \var{NOx} := $\var{throttle}^3\mathbin{/}(\const{k}\cdot\var{def\_dose})$
    \EndProcedure
  \end{algorithmic}
  \caption{A simple emission control.}\label{fig:ecu:general}
\end{wrapfigure}
Consider the electronic control unit (ECU) of a diesel vehicle, in
particular its exhaust emission control module.  For diesel engines,
the controller injects a certain amount of a specific fluid (an
aqueous urea solution) into the exhaust pipeline in order to lower
mono-nitrogen oxides (\NOx) emissions.  We simplify this control
problem to a minimal toy example. In \cref{fig:ecu:general} we display
a function that reads the \var{throttle} position and calculates which
is the dose of diesel exhaust fluid (DEF) (stored in \var{def\_dose})
that should be injected to reduce the \NOx\ emission.  The last line
of the program precisely models the \NOx\ emission by storing it in
the output variable \var{NOx} after a (made up) calculation directly
depending on the \var{throttle} value and inversely depending on the
\var{def\_dose}.

\begin{wrapfigure}[11]{r}{6cm}
  \algorithmsize\vspace{-2em}  \begin{algorithmic}
    \Procedure{EmissionControl}{{}}
      \State \Call{read}{\var{throttle}}
      \If {$\var{throttle}\in\const{ThrottleTestValues}$}
      \State \var{def\_dose} := \Call{SCRModel}{\var{throttle}}
      \Else
      \State \var{def\_dose} := \Call{altSCRModel}{\var{throttle}}
      \EndIf
      \State \var{NOx} := $\var{throttle}^3\mathbin{/}(\const{k}\cdot\var{def\_dose})$
    \EndProcedure
  \end{algorithmic}
  \caption{A doped emission control.}\label{fig:ecu:doped}
\end{wrapfigure}
The Volkswagen emission scandal arose precisely because their software
was instrumented so that it works as expected
\citeref{https://events.ccc.de/congress/2015/Fahrplan/events/7331.html}        {\emph{only if} operating in or very close to the lab testing conditions}.
For our simplified example, this behaviour is exemplified by the
algorithm of \cref{fig:ecu:doped}. Of course, the real case was less
simplistic.
Precisely, in this setting, the lab conditions define the set of
standard inputs, i.e., the set $\Norm$ is actually
\const{ThrottleTestValues} and, as a consequence, a software like
this one trivially meets the characterisation of \emph{clean} given in
\cref{def:clean:seq}.
However, this unit is intentionally programmed to defy the regulations
when being unobserved and hence it falls directly within our intuition
of what a doped software is (see \cref{dope:intuition}).

The spirit of the emission tests is to verify that the amount of
\NOx\ in the car exhaust gas does not exceed a given threshold
\emph{in general}.  Thus, one would expect that if the input values of
the \textsc{EmissionControl} function deviates within ``reasonable
distance'' from the \emph{standard} input values provided during the
lab emission test, the amount of \NOx\ found in the exhaust gas is
still within the regulated threshold, or at least it does not exceed
it more than a ``reasonable amount''.
A similar rationale could be applied for regulation of other systems
such as
speed limit controllers in scooters and electric bikes.

Therefore, we need to introduce two notions of distance
$d_\Inputs:(\Inputs\times\Inputs)\to\Reals_{\geq0}$ and
$d_\Outputs:(\Outputs\times\Outputs)\to\Reals_{\geq0}$ on inputs and
outputs respectively. In principle, we do not require them to be
metrics, but they need to be commutative and satisfy that
$d_\Inputs(\inp,\inp) = d_\Outputs(\outp,\outp) = 0$ for all
$\inp\in\Inputs$ and $\outp\in\Outputs$.
Since programs are non-deterministic, we need to lift the output
distance to sets of outputs and for that we will use the Hausdorff
lifting which, as we will see, is exactly what we need.
Given a distance $d$, the Hausdorff lifting $\Hausdorff(d)$ is defined by
\begin{equation}\label{def:hausdorff:dist}
  \textstyle
  \Hausdorff(d)(A,B) = \max
  \big\{ \sup_{a\in A}\inf_{b\in B} d(a,b),
  \sup_{b\in B}\inf_{a\in A} d(a,b) \big\}
\end{equation}
Based on this, we provide a new definition that considers two
parameters: parameter $\inpbound$ refers to the acceptable distance an
input may deviate from the norm to be still considered, and parameter
$\outpbound$ that tells how far apart outputs are allowed to be in
case their respective inputs are within $\inpbound$
distance.

\begin{definition}\label{def:ed-clean:seq}
  A parameterised program $S$ is \emph{\robustly\ clean} if for all pairs of
  parameters of interest $\param,\param'\in\ParamofInt$ and inputs
  $\inp,\inp'\in\Inputs$, if $\inp\in\Norm$ is a standard input and
  $d_\Inputs(\inp,\inp')\leq\inpbound$ then
  $\Hausdorff(d_\Outputs)(S(\param)(\inp),S(\param')(\inp')) \leq
  \outpbound$.
                          \end{definition}

Requiring that
$\Hausdorff(d_\Outputs)(S(\param)(\inp),S(\param')(\inp')) \leq
\outpbound$ is equivalent to demand that
\begin{enumerate}
\item\label{def:ed-clean:seq:i}  for all $\outp\in S(\param)(\inp)$ there exists $\outp'\in
  S(\param')(\inp')$ such that $d_\Outputs(\outp,\outp') \leq
  \outpbound$, and
\item\label{def:ed-clean:seq:ii}  for all $\outp'\in S(\param')(\inp')$ there exists $\outp\in
  S(\param)(\inp)$ such that $d_\Outputs(\outp,\outp') \leq
  \outpbound$.
\end{enumerate}
Notice that this is what we actually need for the non-deterministic
case: each output of one of the program instances should be matched
within ``reasonable distance'' by some output of the other program
instance.

Notice that $\inp'$ does not need to satisfy $\Norm$, but it will be
considered as long as it is within $\inpbound$ distance of any input
satisfying $\Norm$.  In such a case, outputs generated by
$S(\param')(\inp')$ will be requested to be within $\outpbound$ distance
of some output generated by the respective execution induced by a
standard input.
In addition, notice that if the program $S$ is
deterministic and terminating we could simply write that
$d_\Outputs(S(\param)(\inp),S(\param')(\inp'))\leq\outpbound$.

The concept of \robustly\ clean programs generalises that of clean programs.
Indeed, by taking $d_\Inputs(\inp,\inp)=0$ and
$d_\Inputs(\inp,\inp')>\inpbound$ for all $\inp\neq \inp'$, and
$d_\Outputs(\outp,\outp)=0$ and $d_\Outputs(\outp,\outp')>\outpbound$
for all $\outp\neq \outp'$, we see that \cref{def:clean:seq} is
subsumed by \cref{def:ed-clean:seq}.
Also, notice that the tolerance parameters $\inpbound$ and $\outpbound$ are
values that should be provided as well as the notions of distance
$d_\Inputs$ and $d_\Outputs$, and, together with the set $\ParamofInt$
of parameters of interest and the set $\Norm$ of standard inputs, are
part of the contract that ensures that the software is \robustly\ clean.
Moreover, the limitation to these tolerance values has to do with the
fact that, beyond it, particular requirements (e.g. safety) may
arise.
For
instance, a smart battery may stop accepting charge if the
current emitted by a standardised but foreign charger is higher than
``reasonable'' (i.e. than the tolerance values); however, it may still
proceed in case it is dealing with a charger of the same brand
for which it may know that it can resort to a customised protocol
allowing ultra-fast charging in a safe manner.

\begin{example}\label{ex:ecu:ed}
We remark that \cref{def:ed-clean:seq} will actually detect as doped
the program of \cref{fig:ecu:doped} for appropriate distances
$d_\Inputs$ and $d_\Outputs$ and tolerance parameters $\inpbound$ and
$\outpbound$. Indeed, suppose that $\textsc{SCRModel}(x)= x^2$,
$\textsc{altSCRModel}(x)= x$, and $\const{k}=2$.  To check if the
programs are \robustly\ clean, take $\Inputs=(0,2]$ (these are the
values that variable \var{throttle} takes), $\Norm=(0,1]$, let the
distances $d_\Inputs$ and $d_\Outputs$ be the absolute values of the
differences of the values that take \var{throttle} and \var{NOx},
respectively, and let $\inpbound=2$ and $\outpbound=1$. With this
setting, the program of \cref{fig:ecu:general} is \robustly\ clean
while the program of \cref{fig:ecu:doped} is not.
\end{example}

\cref{def:ed-clean:seq} can be further generalised by adjusting to a
precise desired granularity given by a function
$f:\Reals\to\Reals\cup\{\infty\}$ that relates the distances of the
input with the distances of the outputs as follows.

\begin{definition}\label{def:f-clean:seq}
  A parameterised program $S$ is \emph{$f$-clean} if for all pairs of
  parameters of interest $\param,\param'\in\ParamofInt$ and inputs
  $\inp,\inp'\in\Inputs$, if $i\in\Norm$ is a standard input then
  $\Hausdorff(d_\Outputs)(S(\param)(\inp),S(\param')(\inp')) \leq
  f(d_\Inputs(\inp,\inp'))$.
                          \end{definition}

Like for \cref{def:ed-clean:seq}, the definition of
$f$-clean does not require $\inp'$ to satisfy $\Norm$.  Moreover, notice
that it is important that $f$ can map into $\infty$, in which case it
means that input $\inp'$ becomes irrelevant to the property.
Also here the Hausdorff distance is elegantly encoding the requirement that
\begin{enumerate}
\item\label{def:f-clean:seq:i}  for all $\outp\in S(\param)(\inp)$ there exists $\outp'\in
  S(\param')(\inp')$ s.t.\ $d_\Outputs(\outp,\outp') \leq
  f(d_\Inputs(\inp,\inp'))$, and
\item\label{def:f-clean:seq:ii}  for all $\outp'\in S(\param')(\inp')$ there exists $\outp\in
  S(\param)(\inp)$ s.t.\ $d_\Outputs(\outp,\outp') \leq
  f(d_\Inputs(\inp,\inp'))$.
\end{enumerate}

This definition is strictly more general than \cref{def:ed-clean:seq},
which can be seen by taking $f$ defined by $f(x)=\outpbound$ whenever
$x\leq\inpbound$ and $f(x)=\infty$ otherwise. (Notice here the use of
$\infty$.) Also, if the program $S$ is deterministic, we could simply
require that $d_\Outputs(S(\param)(\inp),S(\param')(\inp'))\leq
f(d_\Inputs(\inp,\inp'))$.

In this new definition, the bounding function $f$, together with the
distances $d_\Inputs$ and $d_\Outputs$, the set $\ParamofInt$ of
parameters of interest and the set $\Norm$ of standard inputs, are
part of the contract that ensures that the software is $f$-clean.

\begin{example}\label{ex:ecu:f}
For the example of the emission control take the setting as in
\cref{ex:ecu:ed} and let $f(x)=x/2$.  Then the program of
\cref{fig:ecu:general} is $f$-clean while the program of
\cref{fig:ecu:doped} is not.
\end{example}

We remark that the notion of $f$-clean strictly relates the distance
of the input values with the distance of the output values. Thus,
e.g., the accepted distance on the outputs may grow according the
distance of the input grows.  Compare it to the notion of
\robustly\ clean in which the accepted distance on the outputs is only
bounded by a constant ($\outpbound$), regardless of the proximity of
the inputs (which is only observed w.r.t. to constant $\inpbound$).

\section{Software Doping on Reactive Programs}
\label{sec:sd:react}

Though we use the Volkswagen ECU case study as motivation for introducing
\cref{def:ed-clean:seq,def:f-clean:seq}, this program is inherently
reactive: the DEF dosage depends not only of the current inputs but
also on the current state (which in turn is set according to previous
inputs).
Therefore, in this section, we revise the definitions given in the
previous section within the framework of reactive programs.

We consider a parameterised reactive program as a function
$S:\Parameters\to\Inputs^\omega\to 2^{(\Outputs^\omega)}$ so that any
instance of the program reacts to the $k$-th input in the input
sequence producing the $k$-th output in each respective output
sequence.  Thus each instance of the program can be seen, for
instance, as a (non-deterministic) Mealy or Moore machine.
In this setting, we require that $\Norm\subseteq\Inputs^\omega$. Thus,
the definition of a clean reactive program strongly resembles
\cref{def:clean:seq}.

\begin{definition}\label{def:clean:react}
  A parameterised reactive program $S$ is \emph{clean} if for all pairs
  of parameters of interest $\param,\param'\in\ParamofInt$ and input
  $\inp\in\Inputs^\omega$, if $\inp\in\Norm$ then $S(\param)(\inp) =
  S(\param')(\inp)$.
\end{definition}

Naively, we may think that the definition of \robustly\ clean may be also
reused as given in \cref{def:ed-clean:seq} by
considering metrics on $\omega$-traces.  Unfortunately this definition
does not work as expected: suppose two input sequences in $\Inputs^\omega$ that only
differ by a single input in some late $k$-th position but originates a
distance larger than $\inpbound$.  Now the program under study may
become clean even if the respective outputs differ enormously at an
early $k'$-th position ($k'<k$).  Notice that there is no
justification for such early difference on the output, since the input sequences are the same up to position $k'$.

In fact, we notice that the property of being clean is of a safety
nature: if there is a point in a pair of executions in which the program is
detected to be doped, there is no extension of such executions that can
correct it and make the program clean.
In the observation above, the $k'$-th prefix of the trace should be
considered the bad prefix and the program deemed as doped.

Therefore, we consider distances on finite traces:
$d_\Inputs:(\Inputs^*\times\Inputs^*)\to\Reals_{\geq0}$ and
$d_\Outputs:(\Outputs^*\times\Outputs^*)\to\Reals_{\geq0}$.
Now, we provide a definition of \robustly\ clean on reactive programs that
ensures that, as long as all $j$-th prefix of a given input sequence,
with $j\leq k$, are within $\inpbound$ distance, the $k$-th prefix of
the output sequence are within $\outpbound$ distance, for any $k\geq 0$.
In the following definition, we denote with $\inp[..k]$ the $k$-th
prefix of the input sequence $\inp$ (and similarly for output
sequences).

\begin{definition}\label{def:ed-clean:react}
  A parameterised reactive program $S$ is \emph{\robustly\ clean} if for all
  pairs of parameters of interest $\param,\param'\in\ParamofInt$ and
  input sequences $\inp,\inp'\in\Inputs^\omega$, if $\inp\in\Norm$
  then, for all $k\geq 0$ the following must hold
  \[\left(\forall {j\leq k} :
    d_\Inputs(\inp[..j],\inp'[..j])\leq\inpbound\right) \limp
    {\Hausdorff(d_\Outputs)(S(\param)(\inp)[..k],S(\param')(\inp')[..k])
    \leq\outpbound},\]
               where $S(\param)(\inp)[..k]=\{\outp[..k] \mid \outp\in S(\param)(\inp)\}$
  and similarly for $S(\param')(\inp')[..k]$.
                                \end{definition}

By having as precondition that
$d_\Inputs(\inp[..j],\inp'[..j])\leq\inpbound$ \emph{for all} $j\leq k$,
this definition considers the fact that
once one instance of the program deviates too much from the normal
behaviour (i.e.\ beyond $\inpbound$ distance at the input), this
instance is not obliged any longer to meet (within $\outpbound$
distance) the output, even if later inputs get closer again.  This
enables \robustly\ clean programs to stop if an input outside the standard
domain may result harmful for the system.
Also, notice that, by considering the conditions through all $k$-th
prefixes the definition encompasses the safety nature of the
\robustly\ cleanness property.

\medskip\par\noindent\textit{Example~\refstepcounter{example}\label{ex:ecu:react:ed}\ref{ex:ecu:react:ed}.}
A slightly more realistic version of the emission control system on
the ECU is given in \cref{fig:ecu:react}.  It is a closed loop where
the calculation of the DEF dosage also depends on the previous reading
of \NOx.  Moreover, the DEF dosage does not affect deterministically
in the \NOx\ emission.  Instead, there is a margin of error on the
\NOx\ emission which is represented by the factor $\lambda$ and the
non-deterministic assignment of variable \var{NOx} in the penultimate
line within the loop.
\begin{wrapfigure}[11]{r}{6.8cm}
  \algorithmsize\vspace{-.2em}  \begin{algorithmic}
    \Procedure{EmissionControl}{{}}
      \State \var{NOx} := 0
      \Loop
        \State \Call{read}{\var{throttle}}
        \State \var{def\_dose} := \Call{SCRModel}{\var{throttle},\var{NOx}}\vspace{.8ex}
        \State $\var{NOx}\ {} \mathbin{{:}{\in}} \left[(1-\lambda)\frac{\var{throttle}^3}{\const{k}\cdot\var{def\_dose}},(1+\lambda)\frac{\var{throttle}^3}{\const{k}\cdot\var{def\_dose}}\right]$\vspace{.8ex}
        \State \Call{output}{\var{NOx}}
      \EndLoop
    \EndProcedure
  \end{algorithmic}\vspace{-.5em}
  \caption{An emission control (reactive).}\label{fig:ecu:react}
\end{wrapfigure}
This non-deterministic assignment is an (admittedly unrealistic) abstraction 
of the chemical reaction between the exhaust gases and the DEF
dosage.
\cref{fig:ecu:doped:react} gives the version of the emission control
system instrumenting the cheating hack.
We define the selective catalytic reduction (SCR) models as follows:
\[
  \textsc{SCRModel}(x,n)=
  \begin{cases} x^2 & \textrm{if } \const{k}\cdot n\leq x \\
   (1+\lambda)\cdot x^2 & \textrm{otherwise} \\
\end{cases}
\]
where $\lambda=0.1$ and $\const{k}=2$, and $\textsc{altSCRModel}(x,n)=
x$ (i.e., it ignores the feedback of the \NOx\ emission resulting in
the same $\textsc{altSCRModel}$ as in \cref{ex:ecu:ed}). We also take
$\Inputs=(0,2]$ (recall that these are the values that variable
\var{throttle} takes).
The idea of the feedback in $\textsc{SCRModel}$ is that if the
previous emission was higher than expected with the planned current
dosage, then the actual current dosage is an extra $\lambda$ portion
above the planned dosage.

\begin{wrapfigure}[15]{r}{7.1cm}
  \algorithmsize\vspace{-2em}  \begin{algorithmic}
    \Procedure{EmissionControl}{{}}
      \State \var{NOx} := 0
      \Loop
        \State \Call{read}{\var{throttle}}
        \If {$\var{throttle}\in\const{ThrottleTestValues}$}
        \State \var{def\_dose} := \Call{SCRModel}{\var{throttle},\var{NOx}}
        \Else
        \State \var{def\_dose} := \Call{altSCRModel}{\var{throttle},\var{NOx}}
        \EndIf\vspace{.8ex}
        \State $\var{NOx}\ {} \mathbin{{:}{\in}} \left[(1-\lambda)\frac{\var{throttle}^3}{\const{k}\cdot\var{def\_dose}},(1+\lambda)\frac{\var{throttle}^3}{\const{k}\cdot\var{def\_dose}}\right]$\vspace{.8ex}
        \State \Call{output}{\var{NOx}}
      \EndLoop
    \EndProcedure
  \end{algorithmic}
  \caption{A doped emission control (reactive).}\label{fig:ecu:doped:react}
\end{wrapfigure}
For the contract required by \robustly\ cleanness, we let
$\Norm=(0,1]^\omega$ and define
$d_\Inputs(\inp,\inp')=\abs{\last(\inp)-\last(\inp')}$ and similarly
$d_\Outputs(\outp,\outp')=\abs{\last(\outp)-\last(\outp')}$, where
$\last(t)$ is the last element of the finite trace $t$.  We take
$\inpbound=2$ and $\outpbound=1.1$. ($\outpbound$ needs to be a little larger
than in \cref{ex:ecu:ed} due to the non-deterministic assignment to
\var{NOx}.)

In \cref{sec:sd:react:hltl:exp} we will use a model checking tool to prove
that the algorithm in \cref{fig:ecu:react} is \robustly\ clean, while the
algorithm of \cref{fig:ecu:doped:react} is not.
\medskip

As before, \cref{def:ed-clean:react} can be further generalised by
adjusting to a precise desired granularity given by a function
$f:\Reals\to\Reals\cup\{\infty\}$ that relates the distances of the
input with the distances of the outputs as follows.

\begin{definition}\label{def:f-clean:react}
  A parameterised reactive program $S$ is \emph{$f$-clean} if for all
  pairs of parameters of interest $\param,\param'\in\ParamofInt$ and
  input sequences $\inp,\inp'\in\Inputs^\omega$, if $\inp\in\Norm$
  then for all $k\geq 0$,
  $\Hausdorff(d_\Outputs)(S(\param)(\inp)[..k],S(\param')(\inp')[..k])
  \leq f(d_\Inputs(\inp[..k],\inp'[..k]))$.
                                \end{definition}

Like for \cref{def:ed-clean:react}, the definition of $f$-cleanness
also considers distance on prefixes to ensure that major differences
in late inputs do not impact on differences of early outputs,
capturing also the safety nature of the property.

We observe that \cref{def:f-clean:react} is more general than
\cref{def:ed-clean:react}.  As before, define $f$ by
$f(x)=\outpbound$ whenever $x\leq 1$ and $f(x)=\infty$ otherwise, but
also redefine the metric on the input domain as follows:
\[ d_\Inputs^{\text{new}}(\inp[..k],\inp'[..k])=
\begin{cases}
     0 & \text{if } \inp[..k]=\inp'[..k] \\
     1 & \text{if either } \inp\in\Norm \text{ or } \inp'\in\Norm, \inp[..k]\neq \inp'[..k] \\
                & \text{and } d_\Inputs(\inp[..j],\inp'[..j])\leq\inpbound \text{ for all } 0\leq j\leq k  \\
     2 & \text{otherwise}
   \end{cases}
\]
for all $\inp,\inp'\in\Inputs$ and $k\geq0$.

\begin{example}\label{ex:ecu:react:f}
For the example of the emission control take the setting as in
\cref{ex:ecu:react:ed} and let $f(x)=x/2+0.3$. The variation of $f$
w.r.t.\ \cref{ex:ecu:f} is necessary to cope with the non-determinism
introduced in these models.
With this setting, in \cref{sec:sd:react:hltl:exp} we will check that the
program of \cref{fig:ecu:react} is $f$-clean while the program of
\cref{fig:ecu:doped:react} is not.
\end{example}

\section{Analysis through self-composition}
\label{sec:sd:seq:selfcomp}

In this section we will focus on sequential deterministic programs and
we will see them in the usual way: as state transformers.  Thus, if
$\mu,\mu':\Vars\to\Vals$ are states mapping the variables of a program
into values within their domain, we denote with
$(S,\mu)\Downarrow\mu'$ that a program $S$, initially taking values
according to $\mu$, executes and terminates in state $\mu'$.  We
indicate with $(S,\mu)\Downarrow\bot$ that the program $S$ starting at
state $\mu$ does not terminate.  As usual, we denote by
$\mu\models\phi$ that a predicate $\phi$ holds on a state $\mu$.

In this new setting, and restricting to deterministic programs,
\cref{def:clean:seq} could be alternatively formulated as in
\cref{prop:clean:seq:det}.
For this, we will assume that $S$ contains sets of variables
$\vec{x}_\param$, $\vec{x}_\inp$, and $\vec{x}_\outp$ which are
respectively parameter variables, input variables and output
variables.  Moreover, let $\ParamofInt$ and $\Norm$ be predicates on
states containing only program variables in $\vec{x}_\param$ and
$\vec{x}_\inp$, respectively. They characterise the set of parameters
of interest and the set of standard inputs.  Now, we can state,

\begin{proposition}\label{prop:clean:seq:det}
  A sequential and deterministic program $S$ is clean if and only if
  for all states $\mu_1$, $\mu_2$ and $\mu'_1$ such that
  $\mu_1\models\ParamofInt\land\Norm$,
  $\mu_2\models\ParamofInt\land\Norm$,
  $\mu_1(\vec{x}_\inp)=\mu_2(\vec{x}_\inp)$ and
  $(S,\mu_1)\Downarrow\mu'_1$, it holds that
  $(S,\mu_2)\Downarrow\mu'_2$ and
  $\mu'_1(\vec{x}_\outp)=\mu'_2(\vec{x}_\outp)$ for some $\mu'_2$.
\end{proposition}

The proof of the proposition is straightforward since it is basically
a notation change, hence we omit it.  Also, notice that we omit any
explicit reference to non-terminating programs.  This is not necessary
due to the symmetric nature of the predicates.

In the nomenclature of~\cite{BartheDR11:mscs} relations
\begin{align*}
  \mathcal{I} &=\{(\mu_1,\mu_2)\mid{\mu_1\models\ParamofInt\land\Norm,}\\
  &\phantom{=\{(\mu_1,\mu_2)\mid{}}\ \mu_2\models\ParamofInt\land\Norm,
    \text{ and } \mu_1(\vec{x}_\inp)=\mu_2(\vec{x}_\inp)\} \\
  \mathcal{I}' &=\{(\mu_1,\mu_2)\mid {\mu_1(\vec{x}_\outp)=\mu_2(\vec{x}_\outp)}\}
\end{align*}
are called \emph{indistinguishable criteria}\footnote{In this definition, states should actually be considered as
  tuples of values rather than state mappings in order to exactly
  match the definitions of~\cite[Sec.~3]{BartheDR11:mscs}.},
and if $(\mu_1,\mu_2)\in\mathcal{I}$ then we say that $\mu_1$ and
$\mu_2$ are \emph{$\mathcal{I}$-indistinguishable}\footnote{Also, to strictly follow notation
  in~\cite[Sec.~3]{BartheDR11:mscs} we should have written
  $\mu_1\sim^{\mathcal{I}}_{\mathit{id}}\mu_2$ instead of
  $(\mu_1,\mu_2)\in\mathcal{I}$.}.
Similarly, for $\mathcal{I}'$.
Thus, \cref{prop:clean:seq:det} characterises what
in~\cite{BartheDR11:mscs} is called \emph{termination-sensitive
  $(\mathcal{I},\mathcal{I}')$-security} and, by
\cite[Prop.~3]{BartheDR11:mscs}, the property of cleanness can be
analysed using the weakest (conservative) precondition
(\wpre)~\cite{book:dijsktra} through self-composition.

\begin{proposition}
  Let $[\vec{x}/\vec{x}']$ indicate the substitution of each
  variable $x$ by variable $x'$.  Then a deterministic program $S$ is
  clean if and only if
  \begin{gather*}
    \left(\begin{array}{l}
    {(\ParamofInt\land\Norm)} \land
    {(\ParamofInt\land\Norm)[\vec{x}/\vec{x}']} \\[.5ex]
    {} \land {\vec{x}_\inp=\vec{x}'_\inp} \land {\wpre(S,\true)}
    \end{array}\right) \
    \lImp \
    \wpre(S ; S[\vec{x}/\vec{x}'],\vec{x}_\outp=\vec{x}'_\outp).
  \end{gather*}
            \end{proposition}

The term $\wpre(S,\true)$ in the antecedent of the implication is the
weakest precondition that ensures that program $S$ terminates.  It is
necessary in the predicate, otherwise it could become false only
because program $S$ does not terminate.

With the same setting as before, and taking $d_\Inputs$, $d_\Outputs$,
$\inpbound$ and $\outpbound$ as for~\cref{def:ed-clean:seq}, we obtain an
alternative definition of \robustly\ cleanness for deterministic programs.

\begin{proposition}\label{prop:ed-clean:seq:det}
  A sequential and deterministic program $S$ is \robustly\ clean if and only
  if for all states $\mu_1$, $\mu_2$, and $\mu'$ such that
  $\mu_1\models\ParamofInt\land\Norm$, $\mu_2\models\ParamofInt$, and
  $d_\Inputs(\mu_1(\vec{x}_\inp),\mu_2(\vec{x}_\inp))\leq\inpbound$,
  the following two conditions hold:
  \begin{enumerate}
  \item\label{prop:ed-clean:seq:det:i}    if $(S,\mu_1)\Downarrow\mu'$, then $(S,\mu_2)\Downarrow\mu'_2$ and
    $d_\Outputs(\mu'(\vec{x}_\outp),\mu'_2(\vec{x}_\outp))\leq\outpbound$
    for some $\mu'_2$; and
  \item\label{prop:ed-clean:seq:det:ii}    if $(S,\mu_2)\Downarrow\mu'$, then $(S,\mu_1)\Downarrow\mu'_1$ and
    $d_\Outputs(\mu'_1(\vec{x}_\outp),\mu'(\vec{x}_\outp))\leq\outpbound$
    for some $\mu'_1$.
  \end{enumerate}
\end{proposition}

In this case, the indistinguishability criteria are
\begin{align*}
  \!\mathcal{I} &=\{(\mu_1,\mu_2)\mid{\mu_1\models\ParamofInt\land\Norm,         \mu_2\models\ParamofInt,
    \text{ and } d_\Inputs(\mu_1(\vec{x}_\inp),\mu_2(\vec{x}_\inp))\leq\inpbound}\} \\
    \!\!\!\mathcal{I}' &=\{(\mu_1,\mu_2)\mid {d_\Outputs(\mu'_1(\vec{x}_\outp),\mu'_2(\vec{x}_\outp))\leq\outpbound}\}
\end{align*}
Notice that $\mathcal{I}$ is not symmetric.  Then the first item of
\cref{prop:ed-clean:seq:det} characterises termination-sensitive
$(\mathcal{I},\mathcal{I}')$-security while the second item
characterises ter\-mi\-na\-tion-sensitive
$(\mathcal{I}^{-1},\mathcal{I}')$-security.
Using again \cite[Prop.~3]{BartheDR11:mscs}, the property of
\robustly\ cleanness can be analysed using \wpre\ through self-composition.

\begin{proposition}
  A deterministic program $S$ is \robustly\ clean if and only if
  \begin{gather*}
    \ParamofInt \land \Norm \land {\ParamofInt [\vec{x}/\vec{x}']}
    \land {d_\Inputs(\vec{x}_\inp,\vec{x}'_\inp)\leq\inpbound}  \\
    \qquad \lImp \ \left(
    \begin{array}{ll}
    & \wpre(S,\true)
      \lImp
      \wpre(S ; S[\vec{x}/\vec{x}'], d_\Outputs(\vec{x}_\outp,\vec{x}'_\outp)\leq\outpbound) \\[1ex]
    \land \ \
    & \wpre(S[\vec{x}/\vec{x}'],\true)
      \lImp
      \wpre(S[\vec{x}/\vec{x}'] ; S, d_\Outputs(\vec{x}_\outp,\vec{x}'_\outp)\leq\outpbound)
    \end{array}\right)
  \end{gather*}
\end{proposition}

Proceeding in a similar manner, we can also obtain an
alternative definition of $f$-cleanness for deterministic programs.

\begin{proposition}\label{prop:f-clean:seq:det}
  A sequential and deterministic program $S$ is $f$-clean if and
  only if for all states $\mu_1$, $\mu_2$, and $\mu'$ such that
  $\mu_1\models\ParamofInt\land\Norm$, and $\mu_2\models\ParamofInt$,
    the following two conditions hold:
  \begin{enumerate}
  \item\label{prop:f-clean:seq:det:i}    if $(S,\mu_1){\Downarrow}\mu'$, then $(S,\mu_2){\Downarrow}\mu'_2$ and
    $d_\Outputs(\mu'(\vec{x}_\outp),\mu'_2(\vec{x}_\outp))\leq
    f(d_\Inputs(\mu_1(\vec{x}_\inp),\mu_2(\vec{x}_\inp))$ for some
    $\mu'_2$; and
  \item\label{prop:f-clean:seq:det:ii}    if $(S,\mu_2){\Downarrow}\mu'$, then $(S,\mu_1){\Downarrow}\mu'_1$ and
    $d_\Outputs(\mu'_1(\vec{x}_\outp),\mu'(\vec{x}_\outp))\leq
    f(d_\Inputs(\mu_1(\vec{x}_\inp),\mu_2(\vec{x}_\inp))$ for some
    $\mu'_1$.
  \end{enumerate}
\end{proposition}

Notice that the term
$f(d_\Inputs(\mu_1(\vec{x}_\inp),\mu_2(\vec{x}_\inp))$ appears in the
conclusion of the implications of both items.  This may look unexpected
since it seems to be related to the input requirements rather than the
output requirements, in particular because it refers to the input
states.
This makes this case a little less obvious than the previous one. 
To overcome this situation, we introduce a constant
$Y\in\Reals_{\geq0}$ which we assume universally quantified.
Using this, we define the following indistinguishability criteria
\begin{align*}
  \mathcal{I}_Y &=\{(\mu_1,\mu_2)\mid{\mu_1\models\ParamofInt\land\Norm,}\\
    &\phantom{=\{(\mu_1,\mu_2)\mid{}}\
    \mu_2\models\ParamofInt,
    \text{ and } f(d_\Inputs(\mu_1(\vec{x}_\inp),\mu_2(\vec{x}_\inp)))=Y\} \\
  \mathcal{I}'_Y &=\{(\mu_1,\mu_2)\mid {d_\Outputs(\mu'_1(\vec{x}_\outp),\mu'_2(\vec{x}_\outp))\leq Y}\}
\end{align*}

By using this, by \cref{prop:f-clean:seq:det}, we have that $S$ is
$f$-clean if and only if for every $Y\in\Reals_{\geq0}$, and for all
states $\mu_1$, $\mu_2$, and $\mu'$ such that
$(\mu_1,\mu_2)\in\mathcal{I}_Y$
\begin{enumerate}
\item  if $(S,\mu_1)\Downarrow\mu'$, then $(S,\mu_2)\Downarrow\mu'_2$ and
  $(\mu',\mu'_2)\in\mathcal{I}'_Y$ for some $\mu'_2$; and
\item  if $(S,\mu_2)\Downarrow\mu'$, then $(S,\mu_1)\Downarrow\mu'_1$ and
  $(\mu'_1,\mu)\in\mathcal{I}'_Y$ for some $\mu'_1$.
\end{enumerate}

With this new definition, and taking into account again the asymmetry
of $\mathcal{I}_Y$, the first item characterises termination-sensitive
$(\mathcal{I}_Y,\mathcal{I}'_Y)$-security while the second one
characterises termination-sensitive
$(\mathcal{I}_Y^{-1},\mathcal{I}'_Y)$-security.  From this and
\cite[Prop.~3]{BartheDR11:mscs}, the property of $f$-cleanness
can be analysed using \wpre\ and self-composition.

\begin{proposition}\label{prop:f-clean:seq:det:sc}
  A deterministic program $S$ is f-clean if and only if for all $Y\in\Reals_{\geq0}$
  \begin{gather*}
    \ParamofInt \land \Norm \land {\ParamofInt [\vec{x}/\vec{x}']}
    \land {f(d_\inp(\vec{x}_\inp,\vec{x}'_\inp))=Y}  \\
    \qquad \lImp \ \left(
    \begin{array}{ll}
    & \wpre(S,\true)
      \lImp
      \wpre(S ; S[\vec{x}/\vec{x}'], d_\Outputs(\vec{x}_\outp,\vec{x}'_\outp)\leq Y \\[1ex]
    \land \ \
    & \wpre(S[\vec{x}/\vec{x}'],\true)
      \lImp
      \wpre(S[\vec{x}/\vec{x}'] ; S, d_\Outputs(\vec{x}_\outp,\vec{x}'_\outp)\leq Y
    \end{array}\right)
  \end{gather*}
    
\end{proposition}

\begin{figure}[t]
  \begin{align*}
    \wpre(x := e,Q)&=\ Q[e/x] \\
    \hspace{1em}
    \wpre( \algorithmicif\ b\ \algorithmicthen\ S_1\ \algorithmicelse\ S_2\ \algorithmicend\ \algorithmicif,Q) &=\ 
    b \Rightarrow \wpre(S_1,Q) \land\neg b \lImp \wpre(S_2,Q) \\
    \wpre (S_1; S_2,Q) &=\ \wpre (S_1,\wpre(S_2,Q)) \\
    \wpre(\algorithmicwhile\  b\ \algorithmicdo\ S\ \algorithmicend\ \algorithmicdo,Q) &=\ \exists k : {k \geq 0} : H_k(Q)
  \end{align*}
  \hspace{2.5em}
  where
  $H_0(Q) = \neg b \land Q$ and
  $H_{k+1}(Q) = (b \land \wpre(S,H_k (Q))) \vee H_0(Q)$
  \caption{Equations for the \wpre\ calculus}\label{fig:wp}
\end{figure}
\medskip\par\noindent\textit{Example~\refstepcounter{example}\label{ex:ecu:wp}\ref{ex:ecu:wp}.}
In this example, we use \cref{prop:f-clean:seq:det:sc} to prove
correct our statements in \cref{ex:ecu:ed}.  First, we recall the
definition of \wpre\ in \cref{fig:wp}, and rewrite the programs in
\cref{fig:ecu:general,fig:ecu:doped} with all functions and values
properly instantiated in the way we need it here (see
\cref{fig:ecu:general:wp,fig:ecu:doped:wp}).
  
\begin{wrapfigure}[11]{r}{4.2cm}
  \centering\algorithmsize\vspace{-0.3em}  \begin{algorithmic}
    \State \var{def\_dose} := $\var{thrtl}^2$
    \State \var{NOx} := $\var{thrtl}^3\mathbin{/}(2\cdot\var{def\_dose})$
  \end{algorithmic}
  \caption{Program \textsc{ec}.}\label{fig:ecu:general:wp}\par\medskip{~}  \begin{algorithmic}
    \If {$\var{thrtl}\in\const{ThrottleTestValues}$\\\hspace{-.3em}}
    \State \var{def\_dose} := $\var{thrtl}^2$
    \Else
    \State \var{def\_dose} := \var{thrtl}
    \EndIf
    \State \var{NOx} := $\var{thrtl}^3\mathbin{/}(2\cdot\var{def\_dose})$
  \end{algorithmic}
  \caption{Program \textsc{aec}.}\label{fig:ecu:doped:wp}
\end{wrapfigure}
On the one hand, none of the programs have parameters, then $\ParamofInt =
\true$.  On the other hand, $\Norm = (\var{thrtl}\in(0,1])$.  Since
$\wpre(\textsc{ec},\true)=\true$ we have to prove that
\begin{gather}
  \hspace{-1.7em}
  {\var{thrtl}\in(0,1]} \land {\left({\textstyle\frac{\abs{\var{thrtl}-\var{thrtl\,}'}}{2}}=Y\right)}\label{eq:wp:ec}\\
  \hspace{-.5em}
  \lImp \left(\begin{array}{ll}
    & \wpre(\textsc{ec};\textsc{ec}',\abs{\var{NOx}-\var{NOx\,}'}\leq Y)\\[1ex]
    \land & \wpre(\textsc{ec}';\textsc{ec},\abs{\var{NOx}-\var{NOx\,}'}\leq Y)
  \end{array}
  \right)\notag
\end{gather}
where $\textsc{ec}'$ is another instance of \textsc{ec} with every
program variable $x$ renamed by $x'$.  Moreover, function $f$ and
distances $d_\Inputs$ and $d_\Outputs$ are already instantiated.
It is not difficult to verify that
$\wpre(\textsc{ec};\textsc{ec}',{\abs{\var{NOx}{-}\var{NOx\,}'}}\leq Y) \equiv
\left(\frac{\abs{\var{thrtl}{-}\var{thrtl\,}'}}{2}\leq Y\right)$ and
$\wpre(\textsc{ec}';\textsc{ec},{\abs{\var{NOx}{-}\var{NOx\,}'}}\leq Y) \equiv
\left(\frac{\abs{\var{thrtl\,}'{-}\var{thrtl}}}{2}\leq Y\right)$ from
which the implication follows and hence \textsc{ec} is $f$-clean.

For $\textsc{aec}$ we also have that $\wpre(\textsc{aec},\true)=\true$
and hence we have to prove a formula similar to \cref{eq:wp:ec}.  In
this case,
$\wpre(\textsc{aec};\textsc{aec}',{\abs{\var{NOx}-\var{NOx\,}'}}\leq
Y)$ is
\begin{align*}
  & ({\var{thrtl}\in(0,1]} \land {\var{thrtl\,}'\in(0,1]}) \ \lImp \
    \textstyle\frac{\abs{\var{thrtl}-\var{thrtl\,}'}}{2}\leq Y \\
  \land \
  & ({\var{thrtl}\in(0,1]} \land {\var{thrtl\,}'\notin(0,1]}) \ \lImp \
    \textstyle\frac{\abs{\var{thrtl}-\var{thrtl\,}'^2}}{2}\leq Y \\
  \land \
  & ({\var{thrtl}\notin(0,1]} \land {\var{thrtl\,}'\in(0,1]}) \ \lImp \
    \textstyle\frac{\abs{\var{thrtl}^2-\var{thrtl\,}'}}{2}\leq Y \\
  \land \
  & ({\var{thrtl}\notin(0,1]} \land {\var{thrtl\,}'\notin(0,1]}) \ \lImp \
    \textstyle\frac{\abs{\var{thrtl}^2-\var{thrtl\,}'^2}}{2}\leq Y
\end{align*}
The predicate is the same for
$\wpre(\textsc{aec}';\textsc{aec},{\abs{\var{NOx}-\var{NOx\,}'}}\leq Y)$,
since $\abs{a-b}=\abs{b-a}$.
Then, the predicate
\begin{gather*}
  \!\!\!\left({\var{thrtl}\in(0,1]} \land {{\textstyle\frac{\abs{\var{thrtl}-\var{thrtl\,}'}}{2}}=Y}\right)
  \lImp \left(\begin{array}{ll}
    & \wpre(\textsc{aec};\textsc{aec}',\abs{\var{NOx}-\var{NOx\,}'}\leq Y)\\[1ex]
    \land & \wpre(\textsc{aec}';\textsc{aec},\abs{\var{NOx}-\var{NOx\,}'}\leq Y)
  \end{array}
  \right)
\end{gather*}
is equivalent to
\begin{gather*}
  \!\!\!\left({\var{thrtl}\in(0,1]} \land {{\textstyle\frac{\abs{\var{thrtl}-\var{thrtl\,}'}}{2}}=Y}\right)
  \lImp \left(\begin{array}{ll}
    & {\var{thrtl\,}'\in(0,1]} \lImp
    \textstyle\frac{\abs{\var{thrtl}-\var{thrtl\,}'}}{2}\leq Y \\
    \land \
    & {\var{thrtl\,}'\notin(0,1]} \lImp
    \textstyle\frac{\abs{\var{thrtl}-\var{thrtl\,}'^2}}{2}\leq Y
  \end{array}
  \right)
\end{gather*}
which can be proved false if, e.g., $\var{thrtl}=1$ and $\var{thrtl\,}'=1.5$.
\medskip

Notwithstanding the simplicity of the previous example, the technique
can be applied to complex programs including loops.  We decided to
keep it simple as it is not our intention to show the power of
$\wpre$, but the applicability of our definition.

We could profit from \cite{BartheDR11:mscs} for the use of other
verification techniques, including separation logic and model checking
where the properties can be expressed in terms of LTL and CTL.
Particularly, CTL permits the encoding of the full non-deterministic
properties given in \cref{sec:sd:seq}.  We will not dwell on this
since in the next section we explore the encoding of the reactive
properties through a more general setting.

\section{Analysis of reactive programs with HyperLTL}
\label{sec:sd:react:hltl}

HyperLTL~\cite{ClarksonFKMRS14:post} is a temporal logic for the specification
of hyperproperties of reactive systems.  
HyperLTL extends linear-time temporal logic (LTL) with trace quantifiers and trace variables, which allow the logic to refer to multiple traces at the same time. The problem of model checking a HyperLTL
formula over a finite-state model is
decidable~\cite{FinkbeinerRS15:cav}.
In this section, we focus on reactive non-deterministic programs and
use HyperLTL to encode the different definitions of clean reactive
programs given in \cref{sec:sd:react}.
In the following, we interpret a program as a set
$S\subseteq(2^{\ap})^\omega$ of infinite traces over a set 
$\ap$ of atomic propositions.

Let $\pi$ be a \emph{trace variable} from a set $\mathcal{V}$ of trace
variables.  A \emph{HyperLTL formula} is defined by the following
grammar:
\begin{equation}
  \begin{array}{c!{\ {::{=}}\ }c!{\ \mid\ }c!{\ \mid\ }c!{\ \mid\ }c!{\ \mid\ }c}
    \psi & \E{\pi} \psi & \A{\pi} \psi & \multicolumn{3}{l}{\ \ \phi } \\
    \phi & a_\pi & \neg\phi & \phi\lor\phi & \ \X\phi \ \, & \phi\U\phi \\
  \end{array}\label{eq:hyperltl:syntax}
\end{equation}
The quantifiers $\exists$ and $\forall$ quantify existentially and universally, respectively, over the set of 
traces.  For example, the formula
$\A{\pi}\E{\pi'}\phi$ means that for every trace $\pi$ there exists
another trace $\pi'$ such that $\phi$ holds over the pair of traces.
If no universal quantifier occurs in the scope of an existential quantifier, and no existential quantifiers occurs in the scope of a universal quantifier, we call the formula \emph{alternation-free}.
In order to refer to the values of the atomic propositions in the different traces, the atomic propositions are indexed with trace variables:  for some atomic proposition $a\in\ap$ and some trace variable $\pi\in \mathcal{V}$, $a_\pi$
states that $a$ holds in the initial position of
trace $\pi$.  The temporal operators and Boolean connectives are interpreted as usual.
In particular, $\X\phi$ means that $\phi$ holds in the next state of
every trace under consideration. Likewise, $\phi\U\phi'$ means that
$\phi'$ eventually holds in every trace under consideration at the
same point in time, provided $\phi$ holds in every previous instant in
all such traces. We also use the standard derived operators: ${\F\phi}\equiv{\true\U\phi}$,
${\G\phi}\equiv{\neg\F\neg\phi}$, and
${\phi\W\phi'}\equiv{\neg(\neg\phi'\U(\neg\phi\land\neg\phi'))}$.

A \emph{trace assignment} is a partial function
$\tass:\mathcal{V}\to(2^\ap)^\omega$ that assigns traces to variables.
Let $\tass[\pi\mapsto\tr]$ denote the same function as $\tass$ except
that $\pi$ is mapped to the trace $\tr$.  For $k\in\N$, let $t[k]$,
$t[k..]$, and $t[..k]$ denote respectively the $k$-th element of $t$,
the $k$-th suffix of $t$, and the $k$-th prefix of $t$.  The trace
assignment suffix $\tass[k..]$ is defined by
$\tass[k..](\pi)=\tass(\pi)[k..]$.
By $\tass\sat{S}\psi$ we mean that formula $\phi$ is satisfied by the
program $S$ under the trace assignment $\tass$.  Satisfaction is
recursively defined as follows.
\[\begin{array}{l!{\quad \text{iff} \quad}l}
  \tass\sat{S}{\E{\pi} \psi} &
  \tass[\pi\mapsto\tr]\sat{S}\psi \text{ for some } t\in S \\
  \tass\sat{S}{\A{\pi} \psi} &
  \tass[\pi\mapsto\tr]\sat{S}\psi \text{ for every } t\in S \\
  \tass\sat{S}{a_\pi} &
  a\in\tass(\pi)[0] \\
  \tass\sat{S}{\neg\phi} &
  \tass\not\sat{S}\phi \\
  \tass\sat{S}{\phi_1\lor\phi_2} &
  \tass\sat{S}\phi_1 \text{ or } \tass\sat{S}\phi_2 \\
  \tass\sat{S}{\X\phi} &
  \tass[1..]\sat{S}\phi \\
  \tass\sat{S}{\phi_1\U\phi_2} &
  \text{there exists } k\geq 0 \text{ s.t.\ } \tass[k..]\sat{S}\phi_2 \text{ and}\\
  \multicolumn{1}{c}{} &
  \phantom{\text{there exists } i\geq 0 \text{ s.t.\ }}
  \text{for all } 0\leq j < k, \tass[j..]\sat{S}\phi_1
\end{array}\]

We say that a program $S$ \emph{satisfies} a HyperLTL formula $\psi$
if it is satisfied under the empty trace assignment, that is, if
$\emptyset\sat{S}\psi$.

In the following, we give the different characterisations of cleanness
for reactive programs in terms of HyperLTL.
For this, let $\ap = {\ap_\param\cup\ap_\inp\cup\ap_\outp}$ where
$\ap_\param$, $\ap_\inp$, and $\ap_\outp$ are the atomic propositions
that define the parameter values, the input values, and the output
values respectively.
Thus, we take $\Parameters=2^{\ap_\param}$, $\Inputs=2^{\ap_\inp}$ and
$\Outputs=2^{\ap_\outp}$.
Therefore, a program $S\subseteq(2^\ap)^\omega$ can be seen as a function
$\hat{S}:\Parameters\to\Inputs^\omega\to 2^{(\Outputs^\omega)}$ where
\begin{equation}
  t\in S \quad \text{ if and only if } \quad
  (t\downarrow\ap_\outp)\ \in\ \hat{S}(t[0]\cap\ap_\param)(t\downarrow\ap_\inp),
  \label{eq:prog:tr-to-fn}
\end{equation}
with $t\downarrow A$ defined by $(t\downarrow A)[k]=t[k]\cap A$ for all
$k\in\N$.

For the propositions appearing in the rest of this sections, we will
assume that distances between traces are defined only according to its
last element.  That is, for the distance
$d_\Inputs:(\Inputs^*\times\Inputs^*)\to\Reals_{\geq0}$ there exists a
distance $\hat{d}_\Inputs:(\Inputs\times\Inputs)\to\Reals_{\geq0}$
such that
$d_\Inputs(\inp,\inp')=\hat{d}_\Inputs(\last(\inp),\last(\inp'))$
for every $\inp,\inp'\in\Inputs^*$, and similarly for
$d_\Outputs:(\Outputs^*\times\Outputs^*)\to\Reals_{\geq0}$.
Let us call these type of distances \emph{past-forgetful}.
Moreover, we will need the abbreviations given in
\cref{tb:hyperltl:abbrv} for a clear presentation of the formulas.
\begin{table}
  \caption{Syntactic sugar for comparisons between traces}\label{tb:hyperltl:abbrv}
  \centering\noindent  $\displaystyle\begin{array}{r!{\ \, \text{ iff }\ }l}
  \param_{\pi}=\param_{\pi'} &
  \displaystyle\bigwedge_{a\in\ap_\param} a_{\pi}\liff a_{\pi'} \\[3.5ex]
  \inp_{\pi}=\inp_{\pi'} &
  \displaystyle\bigwedge_{a\in\ap_\inp} a_{\pi}\liff a_{\pi'} \\[3.5ex]
  \outp_{\pi}=\outp_{\pi'} &
  \displaystyle\bigwedge_{a\in\ap_\outp} a_{\pi}\liff a_{\pi'}
  \end{array}$
  \hfill
  $\displaystyle\begin{array}{r!{\ \, \text{ iff }\ }l}
  \hat{d}_\Inputs(\inp_{\pi},\inp_{\pi'})\leq\inpbound &
  \displaystyle\bigvee_{\substack{\inp,\inp'\in\Inputs \\ \hat{d}(\inp,\inp')\leq\inpbound}} {\bigwedge_{a\in\inp}a_{\pi}}\land{\bigwedge_{a\in\inp'}a_{\pi'}}\\[7.5ex]
  \hat{d}_\Outputs(\outp_{\pi},\outp_{\pi'})\leq\outpbound &
  \displaystyle\bigvee_{\substack{\outp,\outp'\in\Outputs \\ \hat{d}(\outp,\outp')\leq\outpbound}} {\bigwedge_{a\in\outp}a_{\pi}}\land{\bigwedge_{a\in\outp'}a_{\pi'}}
  \end{array}$
  \vspace{2ex}\par\noindent  $\displaystyle\begin{array}{r!{\ \, \text{ iff }\ }l}
  \hat{d}_\Outputs(\outp_{\pi},\outp_{\pi'})\leq f(\hat{d}_\Inputs(\inp_{\pi},\inp_{\pi'})) &
  \displaystyle\bigvee_{\substack{\outp,\outp'\in\Outputs, \inp,\inp'\in\Inputs \\ \hat{d}(\outp,\outp')\leq f(\hat{d}(\inp,\inp'))}} {\bigwedge_{a\in\inp}a_{\pi}}\land{\bigwedge_{a\in\inp'}a_{\pi'}}\land{\bigwedge_{a\in\outp}a_{\pi}}\land{\bigwedge_{a\in\outp'}a_{\pi'}}
  \end{array}$ 
\end{table}

The set of parameters of interest $\ParamofInt\subseteq\Parameters$
defines a Boolean formula which we ambiguously call $\ParamofInt$.
Also, we let $\Norm$ be an LTL formula with atomic propositions in
$\ap_\inp$, that is, a formula obtained with the grammar in the second
line of \cref{eq:hyperltl:syntax} where atomic propositions have the
form $a\in\ap_\inp$ (instead of $a_\pi$).  Thus $\Norm$ characterises
the set of all input sequences through an LTL formula.  With
$\Norm_\pi$ we represent the HyperLTL formula that is exactly like
$\Norm$ but where each occurrence of $a\in\ap_\inp$ has been replaced
by $a_\pi$.  Likewise, we let $\ParamofInt_\pi$ represent the Boolean
formula that is exactly like $\ParamofInt$ with each occurrence of
$a\in\ap_\param$ replaced by $a_\pi$.
We are now in conditions to state the characterisation of a clean
program in terms of HyperLTL.

\begin{proposition}\label{prop:hyperltl:clean}  A reactive program $S$ is clean if and only if it satisfies the
  HyperLTL formula
    \begin{align}
    \A{\pi_1}\A{\pi_2}\E{\pi'_2}\ &
    ({\ParamofInt_{\pi_1}}\land{\ParamofInt_{\pi_2}}\land{\Norm_{\pi_1})}  \label{eq:hyperltl:clean}\\
    & {} \limp \left({\param_{\pi_2}=\param_{\pi'_2}}\land\G({\inp_{\pi_1}=\inp_{\pi'_2}}\land{\outp_{\pi_1}=\outp_{\pi'_2}})\right) \notag
  \end{align}
\end{proposition}
As it is given, the formula actually states that
\begin{gather*}
  \forall \param_1 : \forall \param_2 : \forall \inp : 
  \param_1,\param_2\in\ParamofInt \land \inp \in \Norm 
  : \hat{S}(\param_1)(\inp) \subseteq \hat{S}(\param_2)(\inp)
\end{gather*}
Because of the symmetry of this definition (namely, interchanging
$\param_1$ and $\param_2$), this is indeed equivalent to
\cref{def:clean:react}.
Notice that in~\cref{eq:hyperltl:clean}, $\pi_2$ quantifies universally the parameter of the second instance, while $\pi'_2$ represents the existence of the output sequence in such instance.
The proofs of
\cref{prop:hyperltl:clean,prop:hyperltl:ed-clean,prop:hyperltl:f-clean}
follow the same structures. So we only provide the proof of
\cref{prop:hyperltl:ed-clean} which is the most involved.

In fact, \cref{prop:hyperltl:ed-clean} below states the
characterisation of a \robustly\ clean program in terms of two HyperLTL
formulas (or as a single HyperLTL formula by taking the conjunction).

\begin{proposition}\label{prop:hyperltl:ed-clean}  A reactive program $S$ is \robustly\ clean under past-forgetful distances
  $d_\Inputs$ and $d_\Outputs$ if and only if $S$ satisfies the
  following two HyperLTL formulas
    \begin{align}
    & \hspace{-1em} \A{\pi_1}\A{\pi_2}\E{\pi'_2} \label{eq:hyperltl:ed-clean}\\
    & ({\ParamofInt_{\pi_1}}\land{\ParamofInt_{\pi_2}}\land{\Norm_{\pi_1})} \notag \\
    & {} \limp \Big( {\param_{\pi_2}=\param_{\pi'_2}}\land{\G({\inp_{\pi_2}=\inp_{\pi'_2}})}\land 
    \big((\hat{d}_\Outputs(\outp_{\pi_1},\outp_{\pi'_2})\leq\outpbound)\W(\hat{d}_\Inputs(\inp_{\pi_1},\inp_{\pi'_2})>\inpbound) \big)\Big) \notag\\
        & \hspace{-1em} \A{\pi_1}\A{\pi_2}\E{\pi'_1} \notag\\
    & ({\ParamofInt_{\pi_1}}\land{\ParamofInt_{\pi_2}}\land{\Norm_{\pi_1})} \notag \\
    & {} \limp \Big({\param_{\pi_1}=\param_{\pi'_1}}\land{\G({\inp_{\pi_1}=\inp_{\pi'_1}})}\land
    \big((\hat{d}_\Outputs(\outp_{\pi'_1},\outp_{\pi_2})\leq\outpbound)\W(\hat{d}_\Inputs(\inp_{\pi'_1},\inp_{\pi_2})>\inpbound)\big)\Big) \notag
  \end{align}
\end{proposition}

The difference between the first and second formula is subtle, but
reflects the fact that, while the first formula has the universal
quantification on the outputs of the program that takes standard input
and the existential quantification on the program that may deviate,
the second one works in the other way around.  Thus each of the
formulas capture each of the $\sup$-$\inf$ terms in the definition of
Hausdorff distance (see~\cref{def:hausdorff:dist}).  To notice this,
follow the existentially quantified variable ($\pi'_2$ for the first
formula, and $\pi'_1$ for the second one).
Also, the weak until operator $\W$ has exactly the behaviour that we
need to represent the interaction between the distances of inputs and
the distances of outputs.  The semantics of $\phi\W\psi$ is defined by
\begin{equation}\label{eq:sem:weak-until}  t\models\phi\W\psi \quad\text{iff}\quad
  \forall k\geq0: (\forall j\leq k: t[j..]\models\neg\psi) \limp t[k..]\models\phi
\end{equation}

Next, we prove \cref{prop:hyperltl:ed-clean}.
\begin{proof}
  We only prove that the first formula captures the bound on the left $\sup$-$\inf$
  term of the definition of Hausdorff distance (see
  eq.~\cref{def:hausdorff:dist}) in \cref{def:ed-clean:react}.  The
  other condition is proved in the same way and corresponds to the
  other $\sup$-$\inf$ term of the Hausdorff distance.
    Taking into account the semantics of the weak until operator given in
  eq.~\cref{eq:sem:weak-until}, the semantics of HyperLTL in general and
  using abbreviations in \cref{tb:hyperltl:abbrv},
  formula~\cref{eq:hyperltl:ed-clean} is equivalent to the following
  statement
    \begin{gather*}
    \forall t_1\in S : \forall t_2\in S : \exists  t'_2\in S : \\
    \ (t_1\models\ParamofInt \land t_2\models\ParamofInt \land t_1\models\Norm)\\
    \ {} \limp \Big( {(t_2[0]\cap\ap_\param)=(t'_2[0]\cap\ap_\param)}
    \land (\forall j\geq0 :  t_2[j]\cap\ap_\inp = t'_2[j]\cap\ap_\inp) \\
    \qquad\ {} \land
    \forall k\geq0: (\forall j\leq k:{\hat{d}_\Inputs(t_1[j]\cap\ap_\inp,t'_2[j]\cap\ap_\inp)}\leq\inpbound)\\
    \phantom{\qquad\ {} \land \forall k\geq0:\ }
    \limp {\hat{d}_\Outputs(t_1[k]\cap\ap_\outp,t'_2[k]\cap\ap_\outp)}\leq\outpbound \Big)
  \end{gather*}
    By applying some definitions and notation changes, this is equivalent to
    \begin{gather*}
    \forall t_1\in S : \forall t_2\in S : \exists  t'_2\in S : \\
    \ ((t_1[0]\cap\ap_\param)\in\ParamofInt \land (t_2[0]\cap\ap_\param)\in\ParamofInt \land (t_1\downarrow\ap_\inp) \in \Norm)\\
    \ {} \limp \Big( {(t_2[0]\cap\ap_\param)=(t'_2[0]\cap\ap_\param)}
    \land (t_2\downarrow\ap_\inp) = (t'_2\downarrow\ap_\inp) \\
    \qquad\ {} \land
    \forall k\geq0: (\forall j\leq k:{\hat{d}_\Inputs(t_1[j]\cap\ap_\inp,t'_2[j]\cap\ap_\inp)}\leq\inpbound)\\
    \phantom{\qquad\ {} \land \forall k\geq0:\ }
    \limp {\hat{d}_\Outputs(t_1[k]\cap\ap_\outp,t'_2[k]\cap\ap_\outp)}\leq\outpbound \Big)
  \end{gather*}
    which, by logic manipulation, is equivalent to
    \begin{gather*}
    \forall \param_1 : \forall \param_2 : \forall \inp_1 : \forall \inp_2 : \forall \outp_1 : \\
    \Big(\exists t_1\in S : \exists t_2\in S : \\
    \qquad\qquad
    (\param_1=(t_1[0]\cap\ap_\param)\in\ParamofInt) \land (\param_2=(t_2[0]\cap\ap_\param)\in\ParamofInt) \\
    \qquad\qquad
    {} \land \inp_1 = (t_1\downarrow\ap_\inp) \land \inp_2 = (t_2\downarrow\ap_\inp) \land \outp_1 = (t_1\downarrow\ap_\outp) \land \inp_1 \in \Norm \Big)\\
    \ {} \limp \exists \outp_2 : \exists  t'_2\in S : \\
    \qquad\qquad
    \Big( {\param_2=(t'_2[0]\cap\ap_\param)}
    \land {\inp_2 = (t'_2\downarrow\ap_\inp)} \land {\outp_2=(t'_2\downarrow\ap_\outp)} \\ 
    \qquad\qquad
    {} \land
    \forall k\geq0: (\forall j\leq k:{\hat{d}_\Inputs(\inp_1[j],\inp_2[j])}\leq\inpbound) 
    \limp {\hat{d}_\Outputs(\outp_1[k],\outp_2[k])}\leq\outpbound \Big)\\[-1.1cm]
  \end{gather*}
    By~\cref{eq:prog:tr-to-fn} and the fact that distances are
  past-forgetful, the previous equation is equivalent to
    \begin{gather*}
    \forall \param_1 : \forall \param_2 : \forall \inp_1 : \forall \inp_2 : \forall \outp_1 : \\
    \Big(
    \param_1,\param_2\in\ParamofInt \land \inp_1 \in \Norm
    \land
    \forall k\geq0: (\forall j\leq k:{d_\Inputs(\inp_1[..j],\inp_2[..j])}\leq\inpbound) \\
      \qquad\qquad\ {} \land \outp_1\in\hat{S}(\param_1)(\inp_1)  \Big)
    \limp \big(\exists \outp_2\in\hat{S}(\param_2)(\inp_2): {d_\Outputs(\outp_1[..k],\outp_2[..k])}\leq\outpbound \big)
  \end{gather*}
  which in turn corresponds to bounding the left $\sup$-$\inf$ term of
  the Hausdorff distance (see~\cref{def:hausdorff:dist})
  in~\cref{def:ed-clean:react},
  \begin{gather*}
    \forall \param_1 : \forall \param_2 : \forall \inp_1 : \forall \inp_2 : \\
    \big(
    \param_1,\param_2\in\ParamofInt \land \inp_1 \in \Norm
    \land
    \forall k\geq0: (\forall j\leq k:{d_\Inputs(\inp_1[..j],\inp_2[..j])}\leq\inpbound)\big) \\
      \qquad\qquad\qquad\quad\ {} \limp \textstyle
       \big(\sup_{\outp_1\in\hat{S}(\param_1)(\inp_1)}
       \inf_{\outp_2\in\hat{S}(\param_2)(\inp_2)}
           {d_\Outputs(\outp_1[..k],\outp_2[..k])}\big)\leq\outpbound
  \end{gather*}
  thus proving this part of the
  proposition.
  \qed
\end{proof}

Finally, we also give the characterisation of an $f$-clean program in
terms of HyperLTL.

\begin{proposition}\label{prop:hyperltl:f-clean}  A reactive program $S$ is $f$-clean under past-forgetful distances
  $d_\Inputs$ and $d_\Outputs$ if and only if $S$ satisfies the
  following two HyperLTL formulas
    \begin{align}
    & \hspace{-1em}
    \A{\pi_1}\A{\pi_2}\E{\pi'_2} \label{eq:hyperltl:f-clean}\\
    & \ ({\ParamofInt_{\pi_1}}\land{\ParamofInt_{\pi_2}}\land{\Norm_{\pi_1})} \notag\\
    & \ {} \limp \left({\param_{\pi_2}=\param_{\pi'_2}}\land{\G({\inp_{\pi_2}=\inp_{\pi'_2}})}\land{\G\left({\hat{d}_\Outputs(\outp_{\pi_1},\outp_{\pi'_2})}\leq{f(\hat{d}_\Inputs(\inp_{\pi_1},\inp_{\pi'_2}))}\right)}\right) \notag\\
    & \hspace{-1em}
    \A{\pi_1}\A{\pi_2}\E{\pi'_1} \notag\\
    & \ ({\ParamofInt_{\pi_1}}\land{\ParamofInt_{\pi_2}}\land{\Norm_{\pi_1})} \notag\\
    & \ {} \limp \left({\param_{\pi_1}=\param_{\pi'_1}}\land{\G({\inp_{\pi_1}=\inp_{\pi'_1}})}\land{\G\left({\hat{d}_\Outputs(\outp_{\pi'_1},\outp_{\pi_2})}\leq{f(\hat{d}_\Inputs(\inp_{\pi'_1},\inp_{\pi_2}))}\right)}\right) \notag
  \end{align}
\end{proposition}

As before, the difference between the first and second formula is
subtle and can be noticed again by following the existentially
quantified variables in each of the formulas.

We remark that the HyperLTL characterisations presented in
\cref{prop:hyperltl:ed-clean,prop:hyperltl:f-clean} can be extended to
any distance of bounded memory, that is, distances such that
$d(t,t')=d(t[k..],t'[k..])$ for every finite traces $t$ and $t'$ and a
fixed bound $k\in\N$.  The solution proceeds by basically using the
same formulas on an expanded and annotated model (with the
expected exponential blow up w.r.t. to the original one).

\begin{example}\label{ex:ecu:react:hyperltl}
  In our running example of the emission control system (see
  \cref{ex:ecu:react:ed,ex:ecu:react:f}), the property of \robustly\ cleanness
  reduces to checking formula
    \begin{align}
    & \hspace{-1em} \A{\pi_1}\A{\pi_2}\E{\pi'_2}
    \label{eq:hyperltl:ed-clean:ex}\\
    & \ \Norm_{\pi_1}
    \limp \Big({\G({\var{t}_{\pi_2}=\var{t}_{\pi'_2}})}\land 
    \big((\hat{d}_\Outputs(\var{n}_{\pi_1},\var{n}_{\pi'_2})\leq\outpbound)\W(\hat{d}_\Inputs(\var{t}_{\pi_1},\var{t}_{\pi'_2})>\inpbound)\big)\Big)\notag
  \end{align}
    and the obvious symmetric formula.  For readability reasons, we
  shorthandedly write \var{t} for \var{thrtl} and \var{n} for
  \var{NOx}.  Notice that any reference to parameters disappears
  since the emission control system does not have parameters, and the set
  of standard inputs is characterised by the LTL formula $\Norm \equiv
  \G(\var{t}\in(0,1])$.
    Likewise, we can verify that the model of the emission control
  system is $f$-clean through the formula
    \begin{align}
    & \hspace{-1em}
    \A{\pi_1}\A{\pi_2}\E{\pi'_2}
    \label{eq:hyperltl:f-clean:ex}\\
    & \ \Norm_{\pi_1} \limp \Big({\G({\var{t}_{\pi_2}=\var{t}_{\pi'_2}})}\land{}
    \G\big({\hat{d}_\Outputs(\var{n}_{\pi_1},\var{n}_{\pi'_2})}\leq{f(\hat{d}_\Inputs(\var{t}_{\pi_1},\var{t}_{\pi'_2}))}\big)\Big) \notag
  \end{align}
    and the symmetric formula.
\end{example}

  \section{Experimental Results}\label{sec:sd:react:hltl:exp}
          
  We verified the cleanness of the emission control system using the
  HyperLTL model checker
  MCHyper~\cite{FinkbeinerRS15:cav}. The
  input to the model checker is a description of the system as an
  Aiger circuit and a hyperproperty specified as an alternation-free
  HyperLTL formula. Since the HyperLTL formulas from the previous
  section are of the form $\forall \pi_1 \forall \pi_2 \exists \pi_2' \ldots$, and are, hence, not alternation-free, MCHyper cannot
  check these formulas directly. However, it is possible to
  prove or disprove such formulas by 
  strengthening the formulas and their negations manually into
  alternation-free formulas that are accepted by MCHyper.
  
  In order to prove that program \textsc{ec} in
  \cref{fig:ecu:general:wp} is robustly clean, we strengthen formula
  \cref{eq:hyperltl:ed-clean:ex} by substituting $\pi_2$ for the existentially quantified variable $\pi_2'$. The resulting formula is alternation-free :
  \begin{align}
    & \hspace{-1em} \A{\pi_1}\A{\pi_2}
    \ \Norm_{\pi_1}
    \limp 
    \big((\hat{d}_\Outputs(\var{n}_{\pi_1},\var{n}_{\pi_2})\leq\outpbound)\W(\hat{d}_\Inputs(\var{t}_{\pi_1},\var{t}_{\pi_2})>\inpbound)\big)\label{eq:hyperltl:ed-clean:ex:i}
  \end{align}
  MCHyper confirms that program \textsc{ec} satisfies~\cref{eq:hyperltl:ed-clean:ex:i}. The program thus also
  satisfies~\cref{eq:hyperltl:ed-clean:ex}.  Notice that we had
  obtained the same formula if we would have started from the formula
  symmetric to~\cref{eq:hyperltl:ed-clean:ex}.

  To prove that program \textsc{aec} in \cref{fig:ecu:doped:wp} is
  doped with respect to \cref{eq:hyperltl:ed-clean:ex}, we negate~\cref{eq:hyperltl:ed-clean:ex} and obtain
  \begin{align}
    & \hspace{-1em} \E{\pi_1}\E{\pi_2}\A{\pi'_2}
    \notag \\
    & \hspace{-0.5em}
    \neg\Big( \Norm_{\pi_1}
    \ \limp \Big({\G({\var{t}_{\pi_2}=\var{t}_{\pi'_2}})}\land 
    \big((\hat{d}_\Outputs(\var{n}_{\pi_1},\var{n}_{\pi'_2})\leq\outpbound)\W(\hat{d}_\Inputs(\var{t}_{\pi_1},\var{t}_{\pi'_2})>\inpbound)\big)\Big)\Big)\notag
  \end{align}
    This formula is of the form $\E{\pi_1}\E{\pi_2}\A{\pi'_2} \ldots$ and, hence, again not alternation-free. We replace the two existential quantifiers with
  universal quantifiers and restrict the quantification to two specific throttle values,
  $a$ for $\pi_1$ and $b$ for $\pi_2$:
      \refstepcounter{equation}\label{eq:hyperltl:ed-clean:ex:ii}  \begin{align}
    & \hspace{-1em} \A{\pi_1}\A{\pi_2}\A{\pi'_2}\tag{\theequation.a}\label{eq:hyperltl:ed-clean:ex:iia} \\
    & \hspace{-0.5em}
    \G({\var{t}_{\pi_1}=a}\land{\var{t}_{\pi_2}=b}) \ \limp {}
    \notag \\
    & \hspace{-0.5em}
    \neg\Big( \Norm_{\pi_1}
    \ \limp \Big({\G({\var{t}_{\pi_2}=\var{t}_{\pi'_2}})}\land 
    \big((\hat{d}_\Outputs(\var{n}_{\pi_1},\var{n}_{\pi'_2})\leq\outpbound)\W(\hat{d}_\Inputs(\var{t}_{\pi_1},\var{t}_{\pi'_2})>\inpbound)\big)\Big)\Big)\notag
  \end{align}
  This transformation is sound as long as there actually exist traces with throttle values $a$ and $b$. We establish this by checking, separately, that the following existential formula is satisfied:
  \begin{align}\label{eq:hyperltl:ed-clean:ex:ii:guarantee}
    \E{\pi_1}\E{\pi_2}\G({\var{t}_{\pi_1}=a}\land{\var{t}_{\pi_2}=b})
  \end{align}
  MCHyper confirms the satisfaction of both formulas, which proves that \cref{eq:hyperltl:ed-clean:ex} is violated by program \textsc{aec}.
Precisely, the counterexample that shows the violation of
\cref{eq:hyperltl:ed-clean:ex} is any pair of traces $\pi_1$ and
$\pi_2$ that makes $\G({\var{t}_{\pi_1}=a}\land{\var{t}_{\pi_2}=b})$
  true in~\cref{eq:hyperltl:ed-clean:ex:ii:guarantee}.
  We proceed similarly for the formula symmetric to
  \cref{eq:hyperltl:ed-clean:ex} obtaining two formulas just as before
  which are also satisfied by \textsc{aec} and hence the original
  formula is not.
    Also, we follow a similar process to prove that \textsc{ec} is $f$-clean
  but \textsc{aec} is not.

\cref{tb:robust:experimental-results} shows experimental results
obtained with
MCHyper\footnote{\url{https://www.react.uni-saarland.de/tools/mchyper/}}
version 0.91 for the verification of robustly cleanness.  The Aiger
models were constructed by discretizing the values of the throttle and
the \NOx.  We show results from two different models, where the values
of the throttle was discretised in steps of 0.1 units in both models
and the values of the \NOx\ in steps of 0.05 and 0.00625.
All experiments where run on under OS X ``El Capitan'' (10.11.6) on a
MacBook Air with a 1.7GHz Intel Core i5 and 4GB 1333MHz DDR3.
In \cref{tb:robust:experimental-results}, the model size is given
in terms of the number of transitions, while the size of the Aiger circuit encoding the model prepared for the property is given in terms of the number of latches and gates.  The specification checked by MCHyper is the formula
indicated in the property column. Formula
(\ref{eq:hyperltl:ed-clean:ex:ii}.b) is the formula symmetric to
\cref{eq:hyperltl:ed-clean:ex:iia}.  For the throttle values $a$ and
$b$ in formulas \cref{eq:hyperltl:ed-clean:ex:iia} and
(\ref{eq:hyperltl:ed-clean:ex:ii}.b), we chose $b=2$ and let $a$ vary
as specified in the property column.
\cref{tb:f:experimental-results} shows similar experimental results
for the verification of $f$-cleanness.  With
(\ref{eq:hyperltl:ed-clean:ex:i}$'$),
(\ref{eq:hyperltl:ed-clean:ex:iia}$'$), and
(\ref{eq:hyperltl:ed-clean:ex:ii}.b$'$) we indicate the similar
variations to \cref{eq:hyperltl:ed-clean:ex:i},
\cref{eq:hyperltl:ed-clean:ex:iia}, and
(\ref{eq:hyperltl:ed-clean:ex:ii}.b) required to verify
\cref{eq:hyperltl:f-clean:ex}.
Model checking takes less than two seconds for the coarse
discretisation and about two minutes for the fine discretisation.

  \begin{table}
    \centering\setlength\tabcolsep{1ex}    \caption{Experimental results from the verification of robust cleanness of \textsc{ec} and \textsc{aec}\label{tb:robust:experimental-results}}
    \begin{tabular}{|c|c|c|c|c|c|c|}
      \hline
      \multirow{2}{*}{Program} & \NOx & model size & \multicolumn{2}{c|}{circuit size} & \multirow{2}{*}{property} & time \\\cline{3-5}
      & step & \#transitions & \#latches & \#gates & & (sec.) \\\hline
      \multirow{2}{*}{\textsc{ec}}
      &    0.05 &  1436 & 17 &   9749 & \cref{eq:hyperltl:ed-clean:ex:i} &  0.92 \\\cline{2-7}
      & 0.00625 & 60648 & 23 & 505123 & \cref{eq:hyperltl:ed-clean:ex:i} & 22.19 \\\hline
      \multirow{8}{*}{\textsc{aec}}
      & \multirow{4}{*}{0.05} & \multirow{4}{*}{3756} & \multirow{4}{*}{19} & \multirow{4}{*}{27574}
         & \cref{eq:hyperltl:ed-clean:ex:iia} $a=0.1$   & 1.62 \\
      &&&&& (\ref{eq:hyperltl:ed-clean:ex:ii}.b) $a=0.1$ & 1.6 \\
      &&&&& \cref{eq:hyperltl:ed-clean:ex:iia} $a=1$     & 1.68 \\
      &&&&& (\ref{eq:hyperltl:ed-clean:ex:ii}.b) $a=1$   & 1.56  \\\cline{2-7}
      & \multirow{4}{*}{0.00625} & \multirow{4}{*}{175944} & \multirow{4}{*}{25} & \multirow{4}{*}{1623679}
         & \cref{eq:hyperltl:ed-clean:ex:iia} $a=0.1$   & 102.07 \\
      &&&&& (\ref{eq:hyperltl:ed-clean:ex:ii}.b) $a=0.1$ &  96.3 \\
      &&&&& \cref{eq:hyperltl:ed-clean:ex:iia} $a=1$     &  97.67 \\
      &&&&& (\ref{eq:hyperltl:ed-clean:ex:ii}.b) $a=1$   &  92.8 \\\hline
    \end{tabular}
  \end{table}

  \begin{table}
    \centering\setlength\tabcolsep{1ex}    \caption{Experimental results from the verification of $f$-cleanness of \textsc{ec} and \textsc{aec}\label{tb:f:experimental-results}}
    \begin{tabular}{|c|c|c|c|c|c|c|}
      \hline
      \multirow{2}{*}{Program} & \NOx & model size & \multicolumn{2}{c|}{circuit size} & \multirow{2}{*}{property} & time \\\cline{3-5}
      & step & \#transitions & \#latches & \#gates & & (sec.) \\\hline
      \multirow{2}{*}{\textsc{ec}}
      &    0.05 &  1436 & 5 &   9869 & (\ref{eq:hyperltl:ed-clean:ex:i}$'$) &  1.08 \\\cline{2-7}
      & 0.00625 & 60648 & 8 & 505285 & (\ref{eq:hyperltl:ed-clean:ex:i}$'$) & 21.74 \\\hline
      \multirow{8}{*}{\textsc{aec}}
      & \multirow{4}{*}{0.05} & \multirow{4}{*}{3756} & \multirow{4}{*}{6} & \multirow{4}{*}{27708}
         & (\ref{eq:hyperltl:ed-clean:ex:iia}$'$) $a=0.1$   & 1.71 \\
      &&&&& (\ref{eq:hyperltl:ed-clean:ex:ii}.b$'$) $a=0.1$ & 1.72 \\
      &&&&& (\ref{eq:hyperltl:ed-clean:ex:iia}$'$) $a=1$     & 1.72 \\
      &&&&& (\ref{eq:hyperltl:ed-clean:ex:ii}.b$'$) $a=1$   & 1.77 \\\cline{2-7}
      & \multirow{4}{*}{0.00625} & \multirow{4}{*}{175944} & \multirow{4}{*}{9} & \multirow{4}{*}{1623855}
         & (\ref{eq:hyperltl:ed-clean:ex:iia}$'$) $a=0.1$   & 95.29  \\
      &&&&& (\ref{eq:hyperltl:ed-clean:ex:ii}.b$'$) $a=0.1$ &  97.48 \\
      &&&&& (\ref{eq:hyperltl:ed-clean:ex:iia}$'$) $a=1$     & 95.57 \\
      &&&&& (\ref{eq:hyperltl:ed-clean:ex:ii}.b$'$) $a=1$   & 95.5 \\\hline
    \end{tabular}
  \end{table}

\section{A comprehensive characterisation}

If we concretely focus on the contract between the society or
the licensee, and the software manufacturer, we can think in a more
general but precise definition.  It emerges by noticing that there is
a partition on the set of inputs in three sets, each one of them
fulfilling a different role within the contract:
\begin{enumerate}
\item  The set $\Norm$ of \emph{standard inputs}.  For these inputs, the
  program is expected to work exactly as regulated.  It is the case,
  e.g., of the inputs defining the tests for the \NOx\ emission.
  Thus, it is expected that the program complies to
  \cref{def:clean:seq} when provided only with inputs in $\Norm$.
\item  The set $\Commit$ of \emph{committed inputs} such that
  $\Commit\cap\Norm=\emptyset$.  These inputs are expected to be close
  according to a distance to $\Norm$ and are not strictly regulated.
  However, it is expected that the manufacturer commits to respect
  certain bounds on the outputs.  This would correspond to the inputs
  that do not behave exactly like the tests for the \NOx\ emission,
  but yet define ``reasonable behaviour'' of the car on the road.
  The behaviour of the program under this set of inputs can be
  characterised either by \cref{def:ed-clean:seq} or
  \cref{def:f-clean:seq}.
\item  All other inputs are supposed to be anomalous and expected to
  be significantly distant from the standard inputs.  In our emission
  control example, this can occur, e.g., if the car is climbing a steep
  mountain or speeding up in a highway.  In this realm the only
  expectation is that the behaviour of the output is continuous with
  respect to the input.
\end{enumerate}

Bearing this partition in mind, we propose the following general
definition.

\begin{definition}\label{def:clean:general:seq}
  A parameterised program $S$ is \emph{clean} (or \emph{doping-free})
  if for all pairs of parameters of interest
  $\param,\param'\in\ParamofInt$ and inputs $\inp,\inp'\in\Inputs$,
  \begin{enumerate}
  \item\label{def:general:stdin}    if $\inp\in\Norm$ then $S(\param)(\inp) = S(\param')(\inp)$;
  \item\label{def:general:commit}    if $\inp\in\Norm$ and $\inp'\in\Commit$ then
    $\Hausdorff(d_\Outputs)(S(\param)(\inp),S(\param')(\inp')) \leq
    f(d_\Inputs(\inp,\inp'))$.
                                              \item\label{def:general:other}    for every $\epsilon>0$ there exists $\delta>0$ such that for all
    $\inp'\notin\Norm\cup\Commit$ and $\inp\in\Inputs$,
    $d_\Inputs(\inp,\inp')<\delta$ implies
    $\Hausdorff(d_\Outputs)(S(\param)(\inp),S(\param')(\inp'))<\epsilon$.
                                                    \end{enumerate}
\end{definition}

Notice that, while $\ParamofInt$, $\Norm$, $\Commit$, $d_\Inputs$,
$d_\Outputs$, and $f$ are part of the contract entailed by the
definition, $\epsilon$ and $\delta$ in item \ref{def:general:other}
are not since they are quantified (universally and existentially,
resp.)  in the definition.
In this case, we choose for item \ref{def:general:other} to
require that the program $S$ is uniformly continuous in
$\Inputs\setminus(\Norm\cup\Commit)$.  However, we could have opted
for stronger requirements such as Lipschitz continuity.
The chosen type of continuity would also be part of the contract.
Notice that this is the only case in which we require continuity.
Instead, discontinuities are allowed in cases 1 and 2 as long as the
conditions are respected since they may be part of the specification.
In particular, notice that $f$ could be \emph{any} function.
Obviously, a similar definition can be obtained for reactive systems.

We remark that cases~\ref{def:general:stdin}
and~\ref{def:general:commit} can be verified, as we showed in the
paper.  We have not yet explored the verification of
case~\ref{def:general:other}.

\section{Related work}

The term ``software doping'' has being coined by the press about a
year ago and, after the Volkswagen exhaust emissions scandal, the
elephant in the room became unavoidable: software developers introduce
code intended to deceive~\cite{HattonG16:software}.  Recently, a
special session at ISOLA 2016 was devoted to this
topic~\cite{DBLP:conf/isola/2016-2}.  In~\cite{Baum16:isola}, Baum
attacks the problem from a philosophical point of view and elaborates on the ethics of it.  In~\cite{BartheDFH16:isola}, we provided
a first discussion of the problem and some informal characterisations
hinting at the formal proposal of this paper. Though all these works point out the need for a  technical attack on the problem, none of them
provide a formal proposal.

Similar to software doping, backdoored software is a class of software
that does not act in the best interest of users; see for instance the
recent analysis in~\cite{SchneierFKR15}.
The primary emphasis of backdoored software is on leaking confidential
information while guaranteeing functionality.

Dope-freedom in sequential programs is strongly
related to abstract
non-intereference~\cite{BartheDR04:csfw,GiacobazziM04:popl} as already
disussed in \cref{sec:sd:seq:selfcomp}.
More generally, our notions of dope-freedom are hyperproperties~\cite{ClarksonS08}, a
general class that encompasses notions across different domains, in
particular non-interference in security~\cite{TerauchiA05}, robustness
(a.k.a.\, stability) in cyber-physical systems~\cite{ChaudhuriGL10},
and truthfulness in algorithmic game theory~\cite{BartheGGHRS15}.
There exist several methods for verifying hyperproperties, including
relational and Cartesian Hoare logics~\cite{Benton04,Yang07,SousaD16},
self-composition and product programs
constructions~\cite{BartheCK11,BartheDR11:mscs}, temporal logics~\cite{ClarksonFKMRS14:post,FinkbeinerRS15:cav,FinkbeinerH16},
or games~\cite{MilushevC13}. These techniques greatly vary in
their completeness, efficiency, and scalability.

Another worthwhile direction to study is the use of program equivalence
analysis~\cite{GodlinS13,FelsingGKRU14:ase} for the analysis of cleanness.

\section{Concluding remarks}
\label{sec:conclu}

This article has  focused on  a serious and yet long overlooked
problem, arising if software developers intentionally and silently deviate from
the intended objective of the developed software.
A notorious reason behind such deviations are simple and blunt lock-in strategies, so as to
bind the software licensee to a certain product or product family.  
However, the motivations can be more diverse and obscure. As the
software manufacturer has full control over the development process, the 
deviation can be subtle and surreptitiously introduced in a way that
the fact that the program does not quite conform to the expected
requirements may go well unnoticed. 

We have pioneered the formalisation of this problem domain by offering several formal characterisations of software doping. 
These can serve as a framework for establishing a contract between
the interested parties, namely the society or the licensee, and the
software manufacturer, so as to avoid and eventually ban the development of  doped
programs.

We have also reported on the use of existing theories and tools at hand to demonstrate that the formal characterisation can indeed be analysed in various ways. In particular, the application of the self-composition technique opens many research directions for further
analysis of software doping as it has been widely studied in the area
of security~\cite{KovacsSF13,HawblitzelHKLPR15}, semantical
differences~\cite{LahiriHKR12} and cross or relative
verification~\cite{HawblitzelLPHGFDW13}.

As we have demonstrated, the use of HyperLTL enables the automatic analysis of reactive
models with respect to software doping.  However, the complexity of this technique imposes some serious limits on its applicability. Thus, further studies in this direction are needed in order to enable analysis of reactive models of relatively large size,
or alternatively to analyse the program code directly.

We believe our characterisations provide a first solid step to
understand software doping and that our result opens a large umbrella
of new possibilities, both in the direction of more dedicated
characterisations as well as specifically tailored analysis
techniques.  For instance, the idea of dealing with distances and
thresholds already rises the question of whether such distances
could be quantified by probabilities.  Also, the \NOx\ emission
example would immediately suggest that the technique should also be
addressed with testing.  Moreover, the fact that the characterisations
are hyperproperties also invites us to investigate for static analysis
of source code based on type systems, abstraction techniques, etc.

\paragraph{\ackname}
We would like to thank the Dependable Systems and Software Group
(Saarland University) for a fruitful discussion during an early
presentation of this work, and Nicol\'as Wolovick for drawing our
attention to electronic voting.

\begin{sloppypar}
\bibliographystyle{splncs03}

\end{sloppypar}

\end{document}